\documentclass{aa}    
\usepackage[utf8]{inputenc}
\usepackage{rotating} 

\usepackage{array,siunitx}
\usepackage{graphicx,url,twoopt,natbib}
\usepackage[varg]{txfonts}           
\usepackage{hyperref}                
\usepackage{xcolor}
\usepackage{soul}
\hypersetup{
  colorlinks=true,   
  urlcolor=blue,     
  linkcolor=blue,     
  citecolor=blue
}

\bibpunct{(}{)}{;}{a}{}{,}    
\makeatletter
\newcommand{\bibnote}[2]{\global\@namedef{#1note}{#2}}
\newcommand{\biblink}[2]{\global\@namedef{#1link}{#2}}
\makeatother

\newcommand{\msun}{$M_{\odot}$}
\newcommand{\mdot}{$\dot{M}_{\rm acc}$}
\newcommand{\teff}{$T_{\rm eff}$}
\newcommand{\logg}{$\log{g}$}
\newcommand{\vsini}{$\nu\sin{i}$}
\newcommand{\vmic}{$\nu_{\rm mic}$}
\newcommand{\vr}{$\nu_{\rm r}$}

\newcommand{\starone}{HD\,21997}
\newcommand{\startwo}{49\,Ceti}
\newcommand{\starthree}{HD\,131835}
\newcommand{\starfour}{HD\,138813}
\newcommand{\starfive}{HD\,110058}
\newcommand{\starsix}{HD\,32297}
\newcommand{\starcomp}{HD\,100546}
\newcommand{\xhobs}{$\left(\frac{X}{H}\right)_{\rm obs}$}
\newcommand{\xhacc}{$\left(\frac{X}{H}\right)_{\rm acc}$}
\newcommand{\xhref}{$\left(\frac{X}{H}\right)_{\rm ref}$}

\newcommand{\fph}{$f_{\rm ph}$}

\begin{document}  

  \title{Abundance analysis of stars hosting gas-rich debris disks}
  \author{Sandipan P. D. Borthakur\inst{1,2,3}\thanks{E-mail: sandipan.borthakur@ut.ee}, Mihkel Kama\inst{4,1},
  Luca Fossati\inst{2},
  Quentin Kral\inst{5},
  Colin P. Folsom\inst{1},\\
  Johanna Teske\inst{6},
  Anna Aret\inst{1}
  }
  \authorrunning{S. P. D. Borthakur et al.}

    \institute{
        Tartu Observatory, University of Tartu, Observatooriumi 1, T\~{o}ravere, 61602, Estonia
        \and Space Research Institute, Austrian Academy of Sciences, Schmiedlstrasse 6, 8042, Graz, Austria
        \and
        Institute for Theoretical and Computation Physics, Graz University of Technology, Petersgasse 16, 8010 Graz, Austria
        \and Department of Physics and Astronomy, University College London, London, WC1E 6BT, UK
        \and LESIA, Observatoire de Paris, Universit\'e PSL, CNRS, Sorbonne Universit\'e, Universit\'e Paris Cit\'e, 5 place Jules Janssen, F-92195 Meudon, France
        \and Earth and Planets Laboratory, Carnegie Institution for Science, 5241 Broad Branch Road, NW, Washington, DC 20015, USA
    }

    \date{Received October 31, 2024 / accepted February 20, 2025}
    
  \abstract
  {Accretion from protoplanetary or debris disks can contaminate the stellar photosphere, especially in stars that have radiative envelopes. Due to relatively slower photospheric mixing, these stars can exhibit clear contamination signatures. The contaminated photosphere reflects ongoing disk processes, detectable through stellar spectroscopy.
  }
  {We investigate the composition of six gas-rich debris disk-hosting A-type stars to understand possible links with their debris disk or earlier accretion stages.}
  { We used archival spectra and the {\sc zeeman} spectral synthesis code to estimate the stellar parameters and abundances of six debris disk-hosting A-type stars. We also estimated the stellar photospheric accretion contamination parameter, \fph, using {\sc CAMstars} which indicates the fraction of accreting material on the stellar photosphere.}
  {The oxygen abundance in intermediate-mass stars decreases with age until the debris disk stage ($<20\,$Myr), after which it may rise. The downward trend could result from $\rm H_2O$ ice accumulating in dust traps or the formation of hydrated asteroids in the protoplanetary disk, locking oxygen in solids and reducing its accretion onto the star. All stars have similar volatile abundances (C, O), but \starfive\ and \starsix\ show refractory depleted abundances. The near-zero \fph\ values in the six stars suggest that any currently accreted gas would not overwhelm mixing in the photosphere, and would not impact the observed composition. The refractory depleted abundances in \starfive\ and \starsix\ suggest residual or even chronic, accretion contamination from their earlier protoplanetary stages when the accretion rates were about five orders of magnitude higher. For \starfive, with the highest refractory depletion, we estimated a lower limit on its earlier protoplanetary accretion rate of  $9\times 10^{-8}\,M_{\odot}$/yr, similar to other Herbig stars and equal to the Herbig star - \starcomp. This supports our hypothesis that refractory depletion in \starfive\ originates from a prior phase of higher accretion of dust-poor material. We further develop this hypothesis by comparing HD 110058 with the similarly-aged HD 100546 protoplanetary disk system.
  }
  {}
  
  \maketitle

\section{Introduction}
Debris disks are circumstellar dust belts where planetesimal collisions continually replenish the dust. These disks are the next stage of evolution following the dissipation of protoplanetary disks. Several debris disks around early-type stars are known to contain substantial amounts of gas \citep{2013MoorHD21997,2017Moor,2020Kral}. This gas may represent either residual primordial material from the protoplanetary disk \citep{2013KospalHD21997} or secondary gas released from icy planetesimals \citep{2019KralHD131835}. As the debris disk evolves viscously, some of this gas may partially accrete onto the host star \citep{Kraletal2017}. 

The accretion of circumstellar material can cause a detectable composition peculiarity in the photosphere of stars with effective temperatures, \teff\,$\gtrsim6250\,$K (equivalently, masses, $M_{*}\gtrsim1.4\,M_{\rm \odot}$, \citealt{2018JermynCAMStars}). Such stars have mostly radiative envelopes with relatively slower photospheric mixing than convection in low-mass stars. Alternatively, longer-lived composition anomalies may result from the preferential removal of dust from accreting material in the protoplanetary disk stage \citep{2015Kama}. 

Herbig Ae/Be stars are young, intermediate mass ($2$ to $10\,$M$_{\odot}$) stars with an accreting protoplanetary disk \citep{Herbig1960, WatersWaelkens1998}. Following previous studies of protoplanetary disks and stellar abundances, Herbig Ae/Be stars are separated into two distinct groups \citep{2001Meeus}. Stars with gaps in the protoplanetary disks (group\,I disks) have, on average, photospheric abundances depleted in refractory elements compared to stars with no gaps in the disk (group\,II disks, \citealt{2015Kama}). The volatile elemental abundances remain similar in both cases. All elements other than H, He, C, N and O (volatiles) are considered here to be refractory. The ``semi-volatile'' elements, such as Na, S, and Zn with condensation temperatures in the $500$ to $1000\,$K range, are also considered refractories, meaning they are more likely to be locked in dust particles. The refractory depletion of stellar abundances hosting group\,I disk is thought to be caused by dust-poor accretion onto the star due to dust trapping at the outer edge of a disk gap or cavity \citep{2015Kama}. Stars with refractory elemental abundances depleted by a factor of $10-100$, and volatile elemental abundances close to solar values are termed $\lambda\,$Bo\"{o} type peculiar stars \citep{Morganetal1943, Gray1988}.

The accretion rate (\mdot) for Herbig Ae/Be stars is of the order of $10^{-8} - 10^{-6}\,M_{\odot}\,$/yr \citep{2023GrantAccRate}, irrespective of disk gaps. These high accretion rates and relatively slow photospheric mixing cause significant pollution of the stellar photosphere and can leave clear signs in the observed stellar abundances \citep{2012Folsom,2015Kama, 2023GuzmanDiaz}. On the other hand, the observed gas-rich debris disks (disks with CO detection) will also accrete material onto the star, which can have chemical signatures different from those of protoplanetary disks. The calculated gas accretion rates of gas-rich debris disks around A-type stars are on the order of $10^{-13} - 10^{-11} M_{\odot}\,$/yr \citep{Kraletal2017}, which is $3 - 5\,$orders of magnitude lower than that of protoplanetary disks. Due to such low accretion rates, debris disks are unlikely to contaminate the stellar photosphere significantly \citep{2002Turcotte, 2018JermynCAMStars}. However, there has been almost no work seeking abundance anomalies in debris disk host stars. A correlation has been suggested between the $\lambda\,$Bo\"{o} type refractory-poor stars and dusty debris disks \citep{2016Draper,2018Draper}, but no targeted investigation has previously been done of the composition of stars hosting gas-rich debris disks.

In this paper, we analyse the stellar parameters and elemental abundances of six A-type stars with gas-rich debris disks. The goal is to investigate potential signs of accretion contamination from the debris disk or its earlier protoplanetary disk. We also compared the abundances of our sample with Herbig\,Ae/Be stars which host protoplanetary disks. The stellar abundances in this paper are reported as $\log (N_{\rm el}/N_{\rm H})$, unless otherwise stated, where $N_{\rm el}$ and $N_{\rm H}$ are number abundances of a specific element and hydrogen respectively. 

The structure of the article is as follows: Section\,\ref{sec:sample} describes the samples selected for the abundance determination. Section\,\ref{sec:data and analysis} details the spectroscopic data and the methods used to estimate stellar parameters and elemental abundances. Section\,\ref{sec:results} uses the stellar parameters and abundances to analyse accretion contamination on the stellar photosphere from the disk. Section\,\ref{sec:discussion} compares elemental abundances of debris disk-hosting stars with  Herbig\,Ae/Be stars and discusses any link with photospheric contamination from the debris disk or its earlier protoplanetary disk. Section\,\ref{sec:Conclusion} summarises our key findings. 

\section{Sample selection}
\label{sec:sample}

The following considerations guided our sample selection: (1) the stars have a gas-rich debris disk with a high predicted gas accretion rate onto the star (\mdot\,$\gtrsim 10^{-13}\,$\msun\,/yr; \citealt{Kraletal2017}); (2) the stars should have radiative envelopes (\teff\,$\gtrsim$ $6250\,$K), as their relatively slow mixing increases the chance of detectable photospheric ``contamination'' from gas accretion. Any deviations of the stellar surface composition from a normal reference composition may yield insight into past or ongoing accretion. Thus, these deviations provide information about the composition and evolution of the gas disk independently from direct observations. The resulting sample was not exhaustive, but due to data and analysis time limitations, we made the cut at six targets (with the highest gas accretion rates in our sample), with an extended sample to be studied in future work. Below, we briefly describe the six targets, summarising their properties in Table\,\ref{tab:stardiskdata}. 

\begin{table*}[t] 
  \renewcommand{\arraystretch}{1.3}  
  \caption{Details of the six stars used for this study.}  
  \centering
  \begin{tabular}{l   l   l   l   l   l   l   l   l}
  \hline
  \hline
    \multicolumn{1}{c}{Star} & \multicolumn{1}{c}{Sp. Type $\rm ^a$} & \multicolumn{1}{c}{Vmag $\rm ^a$}& \multicolumn{1}{c}{Age}& \multicolumn{1}{c}{$M_{\rm dust}$}& \multicolumn{1}{c}{$M_{\rm CO}$}& \multicolumn{1}{c}{\mdot}& \multicolumn{1}{c}{\fph}& \multicolumn{1}{c}{Age Ref.}\\
     & & & \multicolumn{1}{c}{(Myr)}& \multicolumn{1}{c}{$(M_{\oplus})$}& \multicolumn{1}{c}{$(M_{\oplus})$}&\multicolumn{1}{c}{(\msun/yr)}& & \\
    \hline
    \starone &A3IV/V& $6.374$& $30 - 38$ & $0.27 \pm 0.11$ & $0.050 - 0.85$& $3.56 \times 10^{-11}$ & $5.07 \times 10^{-3}$& 1,2,3\\
    \startwo & A1V& $5.616$& $40 - 50$ & $0.068 \pm 0.014$& $0.061 - 0.35$ & $3.06 \times 10^{-12}$ &$2.62 \times 10^{-5}$& 2,4\\
    \starthree & A2IV& $7.900$& $10 - 16$ & $0.71^{+0.03}_{-0.05}$ & $0.045 \pm 0.015$& $1.07 \times 10^{-11}$ & $3.15 \times 10^{-4}$&2,4,5,6,7\\
    \starfour & A0V& $7.316$& $7 - 13$ & $0.0083 \pm 0.0015$& $0.0001 - 0.003$ & $2.04 \times 10^{-11}$
    &$4.13 \times 10^{-4}$& 4,8\\
    \starfive & A0V& $7.800$& $10 - 17$ & $0.0028 \pm 0.0002$& $0.069^{+3.56}_{-0.057}$ & $1.42 \times 10^{-13}$ &$1.42 \times 10^{-6}$&2,4,5,6,7,8\\
    \starsix & A0V & $8.500$& $10 - 20$ & $0.57 \pm 0.05$ & $0.316^{+0.126}_{-0.190}$ & $1.08 \times 10^{-12}$ & $8.79 \times 10^{-5}$&2,8\\
    \hline\\
  \end{tabular}
  \tablefoot{The \fph\ values are calculated in this paper and are described in Sect.\,\ref{subsec:fph description}.}
  \tablebib{(a) \citet{1993Turcon_sptype_vmag}, (1) \citet{2016Galicher_age}, (2) \citet{2017Meshkat_age}, (3) \citet{2020Ujjwal_age}, (4) \citet{2022Pearce_age}, (5) \citet{2007Rhee_age}, (6) \citet{2019Nielsen_age}, (7) \citet{2011Tetzlaff_age}, (8) \citet{2015Mittal_age}.}
  \label{tab:stardiskdata}
\end{table*}

\subsection{\starone}

\starone\ is an A3IV/V type debris-disk hosting star at a distance of $69.6\,$pc \citep{2022AndersHD21997} belonging to the Columba association \citep{2013MoorHD21997}. The dust and CO gas mass of the disk is $0.27 \pm 0.11\,M_{\oplus}$ \citep{2013MoorHD21997}, and $0.05 - 0.85\,M_{\oplus}$ \citep{2023Cataldi,2020HiguchiHD21997_49Ceti}, respectively. The CO gas disk has an inner radius of $<26\,$au and an outer radius of $138 \pm 20\,$au \citep{2013KospalHD21997}. The dust disk has an inner radius of $55 \pm 16\,$au and an outer radius of $>490\,$au \citep[unconstrained,][]{2013MoorHD21997}. The gas and dust disks are not spatially co-located. The dust is of secondary origin, produced through planetesimal collisions \citep{2013MoorHD21997}. Whether the gas is primordial or secondary is still debated \citep{2023Cataldi}. Neutral C has also been detected in the disk \citep{2023Cataldi}.

\subsection{\startwo}

\startwo\ (HD 9672) is an A1V type debris-disk hosting star at a distance of $58.3\,$pc belonging to the Argus moving group \citep{2012Zuckerman49Ceti}. It has a nearly edge-on disk with a CO gas mass of $0.061 - 0.35\, M_{\oplus}$ \citep{2023Cataldi}. The millimetre (mm) dust grains are distributed from $100 - 310\,$au, peaking at $\sim110\,$au \citep{2017Hughes49Ceti}. The dust mass of the disk is $0.068 \pm 0.014\,M_{\oplus}$ \citep{2017Hughes49Ceti}. The CO gas is distributed from  $30 - 195\,$au \citep{2019Higuchi49Ceti}. The central cavity is depleted in both dust and CO gas. The unusually high C/CO ratio in the disk suggests a secondary origin of CO gas \citep{2017Higuchi49Ceti}. Neutral C and O have also been detected in the disk \citep{2019Higuchi49Ceti,2014Roberge49Ceti}. A transient Falling Evaporating Bodies (FEB)-like event has been detected, possibly due to perturbations by a planetary companion in the inner disk \citep{2012Montgomery49Ceti}.

\subsection{\starthree}

\starthree\ is an A2IV type debris-disk hosting star at a distance of $130.3\,$pc belonging to the Upper Centaurus Lupus sub-association of the Scorpius\,–\,Centaurus (Sco-Cen) association \citep{2015Jang-CondellScoCenMember}. It is a young, nearly edge-on debris-disk hosting star. The dust and CO gas mass of the disk is $0.71^{+0.03}_{-0.05}$ and $0.045 \pm 0.015\,M_{\oplus}$, respectively \citep{2019KralHD131835}. The mm-dust disk belt radius is $90^{+3.8}_{-3.7}\,$au with most of the mass within $50$ and $140$\,au, albeit with large uncertainties \citep{2019KralHD131835}. The observations of micron-sized dust grains from the Spectro-Polarimetric High-contrast Exoplanet REsearch (SPHERE) instrument have revealed three concentric rings \citep{2017FeldtHD131835}. The outermost ring has an inner and outer edge of $85$ and $120\,$au with a peak at $98\,$au. The middle ring has an inner and outer radius of $57$ and $71\,$au, respectively, peaking at $66\,$au. The innermost ring is located at $35 - 45\,$au. Although a planetary companion can produce these rings, other methods \citep[e.g. clumping instability, magnetised disk, etc;][]{2017FeldtHD131835} can induce such structures too. No companion has been detected so far.  The micron-sized dust grains are distributed over a larger range in radii compared to the more compact distribution of mm dust. The CO gas disk has inner and outer radii of $50$ and $250\,$au with a peak emissivity at $\sim100\,$au \citep{2019HalesHD131835_HD138813}. The position of the gas ring matches the positions of the outermost dust rings. Neutral C has also been detected in the disk \citep{2019KralHD131835}. The origin of the CO gas is considered to be secondary \citep{2019KralHD131835}.

\subsection{\starfour}

\starfour\ is a young A0V debris-disk hosting star at a distance of $130.3\,$pc belonging to the Upper Scorpius sub-association of the Sco-Cen association \citep{2015Jang-CondellScoCenMember}. The dust ring is located within $67^{+20}_{-19}$ and $150^{+60}_{-50}\,$au \citep{2016SifryHD138813HD110058ALMA}. The CO gas disk extends till $\sim$181 au, although most of the gas is present within $25-104\,$au \citep{2019HalesHD131835_HD138813}. The dust mass of the disk is $0.0083 \pm 0.0015\,M_{\oplus}$ \citep{2016SifryHD138813HD110058ALMA}. The mass of the CO gas in the disk is $0.0001 - 0.003\,M_{\oplus}$ \citep{2019HalesHD131835_HD138813}.

\subsection{\starfive}
\starfive\ is an A0V type star at a distance of $129.9\,$pc belonging to the Lower Centaurus Crux sub-association of the Sco-Cen association \citep{2015Jang-CondellScoCenMember}. It has an edge-on debris disk \citep{2015KasperHD110058ScatterLight}. There are two different dust mass estimates of the disk $0.080^{+0.002}_{-0.003}\,M_{\oplus}$ \citep{2022HalesALMAHD110058} and $0.0028 \pm 0.0002$ \citep{2016SifryHD138813HD110058ALMA}. The CO gas mass of the disk is $0.069^{+3.56}_{-0.057} \,M_{\oplus}$ \citep{2022HalesALMAHD110058}. The dust disk has an inner and outer radius of $18\,$au ($< 23\,$au with 99.7\% confidence) and $67 \pm 4\,$au, respectively and a dust peak (planetesimal belt) at $31^{+10}_{-8}\,$au \citep{2022HalesALMAHD110058, Stasevic2023HD110058}. The CO gas disk spans $10-30\,$au with a peak radius at $\sim10-20\,$au \citep{2022HalesALMAHD110058}. The gas is expected to be of secondary origin formed through collisions of planetesimals. The inner disk is tilted compared to the outer disk, suggesting perturbation by a massive planet in an inclined orbit in the inner disk \citep{Stasevic2023HD110058}. The CO gas distribution in the inner cavity can either be explained by low viscosity or a massive planet carving the inner cavity and accreting the inflowing gas \citep{2022HalesALMAHD110058}.

\subsection{\starsix}

\starsix\ is an A0V type edge-on debris-disk hosting star at a distance of $132.4\,$pc. The mm-dust has a two-component structure, namely i) a planetesimal belt between $78.5 \pm 8.1\,$au and $122 \pm 3\,$au, and ii) an outer halo from $122 \pm 3\,$au to $440 \pm 32\,$au \citep{2018MacGregorHD32297}. The outer halo consists of smaller dust grains \citep{2022OlofssonHD32297}. The mm-sized dust mass in the belt is $0.57 \pm 0.05\,M_{\oplus}$ and the micron-sized dust mass in the halo is $0.29 \pm 0.06\,M_{\oplus}$ \citep{2018MacGregorHD32297}. The CO gas mass in the disk is $0.316^{+0.126}_{-0.190}\,M_{\oplus}$ \citep{2023Cataldi}. Although mm-dust and gas images show no asymmetry, observations of micron-sized dust grains reveal an asymmetric disk with a swept-back wing-like structure \citep{2014SchneiderHD32297}, which extends to $\sim1560\,$au. The disk has an asymmetric concavity at larger radii. These complex structures in the disks are poorly understood, and their origin can be affected by interaction with a dense interstellar medium (ISM) and/or intrinsic perturbations. The CO gas is co-located with the dust continuum. The observations from the Atacama Large Millimeter/submillimeter Array (ALMA) have detected a neutral C ring at $\sim110\,$au with a full-width half maximum (FWHM) of $\sim80\,$au \citep{2020CataldiHD32297}, slightly extended compared to the planetesimal belt. There has been the detection of Na absorption from the circumstellar disk and slight variability in the radial velocity too \citep{2007RedfieldHd32297}. Such absorption can be due to FEBs or exocomets.

\section{Data and analysis}
\label{sec:data and analysis}
We retrieved archival spectra for each of the six stars. The spectra of \startwo\ were observed using the Echelle SpectroPolarimetric Device
for the Observation of Stars (ESPaDOnS) attached to the Canada France Hawaii Telescope (CFHT) with a spectral resolution of $\sim65\,000$. The data is available on the Polarbase website\footnote{\texttt{\href{http://polarbase.irap.omp.eu/}{http://polarbase.irap.omp.eu/}}}. The spectra for \starone, \starthree, \starfour, \starfive\ and \starsix\ were observed using the Fiber-fed Extended Range Optical Spectrograph (FEROS) mounted on the Max-Planck-Gesellschaft (MPG) 2.2m telescope in La Silla, Chile, with a spectral resolution $\sim48\,000$. The data is available on the European Southern Observatory (ESO) Archive Science Portal\footnote{\texttt{\href{https://archive.eso.org/scienceportal/home}{https://archive.eso.org/scienceportal/home}}}. Multiple spectra from the same instruments were combined for each star to improve the signal-to-noise (S/N) ratio. Details about the analysed spectra are listed in Table\,\ref{tab:stellar_spectra}.


The stellar temperature (\teff) and surface gravity (\logg) were first estimated using photometric calibrations. Using these estimates as initial values, we derived the final stellar parameters and elemental abundances using spectral fitting. Finally, we tested the quality of our \teff\ and \logg\ estimates by comparing theoretical and observed colours. Below, we list a detailed description of the analysis.

\subsection{Photometric Calibration}
As a first estimate for effective temperature (\teff) and surface gravity (\logg), we use photometric colours from \citet{1997Mermilliod}\footnote{\texttt{\href{https://gcpd.physics.muni.cz/}{https://gcpd.physics.muni.cz/}}} and the {\sc TempLogG} \citep{2006Kaiser} software, which estimates stellar parameters from empirical photometric calibrations. The photometric systems and the calibrations used are listed in Table\,\ref{tab:photometric calibration}. All stars except \starsix\ had photometric colours in at least one of the three photometric systems (Str\"{o}mgren, Johnson or Geneva).

\begin{table*}[t]
    \renewcommand{\arraystretch}{1.1}
    \caption{\teff\ and \logg\ estimates from photometric calibration.}
    \newcolumntype{M}{>{\centering\arraybackslash}m{0.6cm}}
    \centering
    \begin{tabular}{c|c|c   c|c   c|c   c|c   c|c   c}
    \hline
    \hline
    \multicolumn{1}{c|}{} & \multicolumn{1}{c|}{}& \multicolumn{2}{c|}{\starone} & \multicolumn{2}{c|}{\startwo} & \multicolumn{2}{c|}{\starthree} & \multicolumn{2}{c|}{\starfour} & \multicolumn{2}{c}{\starfive}\\ 
    \hline
    \multicolumn{1}{c|}{Photometric system} & \multicolumn{1}{c|}{Calibration}  & \teff &\logg& \teff &\logg& \teff &\logg& \teff &\logg& \teff &\logg\\ 
    \hline 
    \multicolumn{1}{r|}{Str\"{o}mgren}& \multicolumn{1}{l|}{\citet{1985Moon}}&8570 &4.39&9106 &4.29&-&-&9681&4.66&-&-\\
     & \multicolumn{1}{l|}{\citet{1993Napiwotzki}}&8411 &4.24&8668 &4.08&-& -&9073&4.19& -&-\\
     & \multicolumn{1}{l|}{\citet{1994Balona}}&8403 &4.40&8827 &4.30& -& - &9724&5.01& -&-\\
     & \multicolumn{1}{l|}{\citet{1997Ribas}}&-&-&9102 &4.29& -&-&9684&4.70&-&-\\
     & \multicolumn{1}{l|}{\citet{1997Castelli}}&-&-&8851 &4.09& -&-&9286&4.39&-&-\\
    \hline
    \multicolumn{1}{r|}{Geneva}& \multicolumn{1}{l|}{\citet{1997Geneva}}&8464&4.25&-&-&- &-& 9060&4.12&-&-\\
    \hline
    \multicolumn{1}{r|}{Johnson}&\multicolumn{1}{l|}{\citet{1998Johnson}}&-&-&-&-& 7716& 4.21&8752&4.23&7823& 4.21\\ \hline
    \end{tabular}
  \label{tab:photometric calibration}
\end{table*}

\subsection{Spectrum fitting}
The {\sc zeeman} spectral synthesis code \citep{1988Landstreet,2001Wadeetal,2012Folsom} was used for spectral fitting along with {\sc atlas9} \citep{1993Kurucz,2004castelliKurucz} stellar model atmospheres and a list of atomic line data from the Vienna Atomic Line Database\footnote{\texttt{\href{https://vald.astro.uu.se/}{https://vald.astro.uu.se/}}} \citep[VALD;][]{1995Piskunovetal,1997Ryabchikovaetal,1999Kupka,2000Kupkaetal,2015Ryabchikova}. {\sc zeeman} is a Fortran-based spectral synthesis code which estimates stellar parameters and elemental abundances by fitting a continuum normalised spectrum using $\chi^2$-minimisation. The code assumes Local Thermodynamic Equilibrium (LTE) in the stellar atmosphere. The spectra were continuum normalised using {\sc suppnet}\footnote{\texttt{\href{https://github.com/RozanskiT/suppnet}{https://github.com/RozanskiT/suppnet}}} \citep{2022suppnet}. {\sc suppnet} is a Python-based neural network code approximating the continuum as a spline. The interactive interface allows the user to tweak the continuum to their preferences. Using the continuum normalised spectrum, the stellar parameters -- \teff, \logg, projected rotational velocity (\vsini), microturbulence velocity (\vmic), radial velocity (\vr), and the stellar elemental abundances were derived using a multi-step procedure. The \teff\ and \logg\ values were initialised either using the photometric estimates from the previous section (Table\,\ref{tab:photometric calibration}) or a typical value of $8000\,$K and $4.0\,$dex (for \starsix), respectively. The \vr\ and \vsini\ values were initialised using rough estimates from manually comparing synthetic spectra with the observations. The initial values of the stellar abundances were assumed to be solar. We initialised the \vmic\ value using the formula from \citet{Pace2006vmic}
\begin{equation}
    \nu_{\rm mic} = -4.7\log\left(T_{\rm eff} \right) + 20.9\, \rm km\ s^{-1}\,.
\end{equation}

The spectrum was first divided into multiple smaller wavelength regions. For \starone, we used the spectral regions: $4169-4606$ (excluding the H$\gamma$ region), $4606-5187$ (excluding the H$\beta$ region), $5187-5387$ and $5387-5672.01\,\AA$. For \startwo, \starthree, \starfour, \starfive\ and \starsix, we used the spectral regions: $4439-4760$, $4976-5176$, $5176-5376$ and	$5376-5673\,\AA$. In our first run, we did not use the spectrum beyond $\sim 5600\,\AA$, which has wiggles in the continuum, possibly due to bad order merging, making it difficult to normalise the spectrum. This was true for all FEROS spectra used in our analysis. To maintain a similar analysis, we also used the same method for the ESPaDOnS spectrum (\startwo). We fit each spectral region independently for the stellar parameters (\teff, \logg, \vsini, \vmic\ and \vr) and abundances. We fit each stellar parameter individually. After estimating the stellar parameters, we fit together the abundances of elements showing the strongest and the largest number of absorption lines in the spectral regions (e.g. Fe, Ti, Cr). One iteration occurs when one round of the fit for all the parameters and the abundances is done. We iterate until all parameters and abundances converge, which typically happens with $3-4$ iterations. The uncertainties of the stellar parameters and the abundances measured in this run were based on the standard deviation of their values obtained from these different wavelength regions. 

Using the estimated stellar parameters, we re-normalised the spectrum, including wavelength regions beyond $5600\,\AA$. During the continuum normalisation, we compared our normalisation with a synthetic normalised spectrum. The synthetic spectrum was produced using {\sc zeeman} and the stellar parameters and the abundances from the first run. This allowed us to remove complex wiggles in the continuum without significantly affecting the line depths of absorption lines. We fixed the stellar parameters and the abundances of elements measured from the first run. In this second run, we estimated the abundances of all other elements (e.g. Sc, C, O), which were not estimated in the first run. Since most lines are blended, to determine the abundance of an element from a spectral line, we fit the spectral region around that line while fixing the abundances of the other elements. These spectral regions are $\lesssim10\AA$, and are much smaller than the spectral regions fit in the first run. Their uncertainties were based on a standard deviation of their line-by-line estimates. For elements with abundance measurements from a single line, we assigned a standard uncertainty of 0.20 dex.

\subsection{Colours}
To check the reliability of our \teff\ and \logg\ estimates, we compared the synthetic colours to the observed ones. The Strömgren colours $(b - y)$ and $c\rm1$ are particularly sensitive to \teff\ and \logg, respectively. A difference of less than 0.05 between the synthetic and observed colours corresponds to estimated uncertainties of $50\,$K in \teff\ and $0.05\,$dex in \logg. The synthetic colours were derived from combining spectral energy distributions computed by {\sc LLmodels} \citep{2004Shulyakllmodels} using the atmospheric parameters obtained from spectroscopy with filter transmission curves. We did not account for measurement errors in the stellar parameters and abundances when calculating the synthetic colours.
The observed colours were only available for \starone, \startwo, and \starfour, which were corrected for interstellar reddening. 

\begin{figure}
\resizebox{\hsize}{!}{\includegraphics[trim=0.1cm 0.5cm 1.0cm 0.5cm]{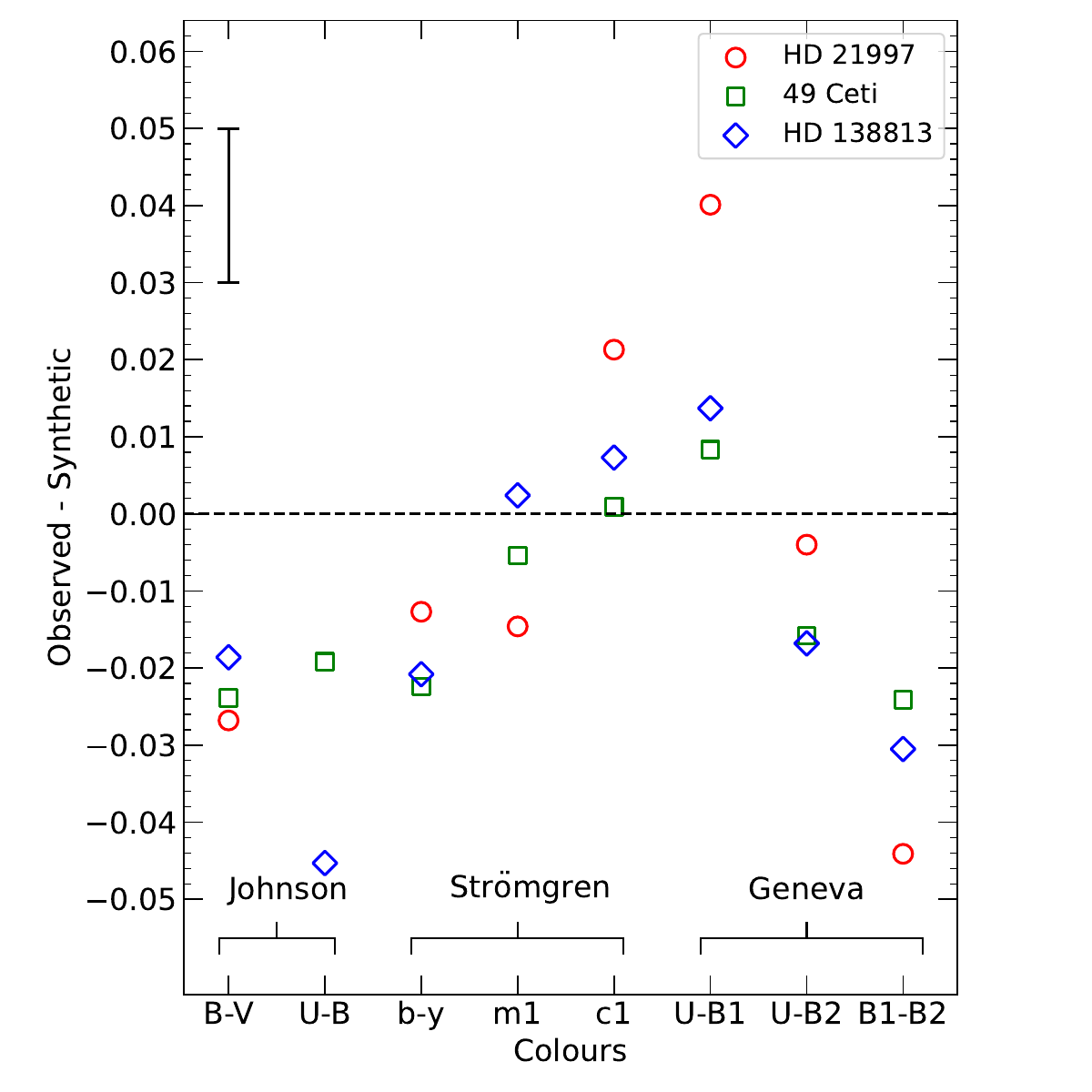}}
 \caption{Observed - Synthetic colours for different photometric systems. Better \teff\ and \logg\ estimates produce a difference closer to zero. The top left corner includes a typical uncertainty in observed colours. For colour difference for \starfour\ is based on $E(B-V) = 0.04$.}
 \label{fig:colour_plot}
\end{figure}

We used the publicly available NASA Galactic Dust Reddening and Extinction map\footnote{\texttt{\href{https://irsa.ipac.caltech.edu/applications/DUST/}{https://irsa.ipac.caltech.edu/applications/DUST/}}}, which is based on the data from \citet{SF2011} and \citet{SFD1998}. Table\,\ref{tab:extinction_table} lists the $E(B-V)$ values for \starone\ and \startwo. The $E(B-V)$ for \starfour\ is reported as $0.2935 \pm 0.0315$ \citep{SF2011} and	$0.3413 \pm 0.0366$ \citep{SFD1998}. Applying a $E(B-V)$ of $0.30$ on the synthetic colours of \starfour, gives a difference between synthetic and observed colours of $>$\,$0.20$. To resolve this issue, we compared the flux-calibrated X-shooter spectrum available in the ESO archive science portal with a flux-calibrated synthetic spectrum produced using {\sc LLmodels} \citep{2004Shulyakllmodels} based on our \teff\ and \logg\ estimates. Based on this comparison, we found $E(B-V) = 0.04$ for \starfour. The discrepancy in the $E(B-V)$ value likely arises because the dust map does not consider the distance to the star. As a result, the dust cloud may be located behind the star, leading to an overestimation of extinction.
Figure\,\ref{fig:colour_plot} shows the difference between synthetic and observed colours for \starone, \startwo\ and \starfour\ where we used the revised $E(B-V)$ of 0.04 for \starfour.
\begin{table}[t]
  \renewcommand{\arraystretch}{1.1}  
  \caption{Extinction parameters used to correct the observed colours.}
  \centering
  \begin{tabular}{l   l   l}
  \hline
  \hline
    \multicolumn{1}{c}{Star} & \multicolumn{1}{c}{$E(B-V)_{\rm SF2011}$} & \multicolumn{1}{c}{$E(B-V)_{\rm SFD1998}$} \\
    \hline
    \starone & $0.0127\pm0.0011$ & $0.0148\pm0.0013$\\
    \startwo & $0.0161\pm0.0003$ &	$0.0187\pm0.0004$\\
    \hline\\
  \end{tabular}
  \tablebib{SF2011: \citet{SF2011}, SFD1998: \citet{SFD1998}}
  \label{tab:extinction_table}
\end{table}

\section{Results} \label{sec:results}
\begin{figure}
\resizebox{\hsize}{!}{\includegraphics[trim=0.0cm 0.2cm 0.0cm 0.0cm]{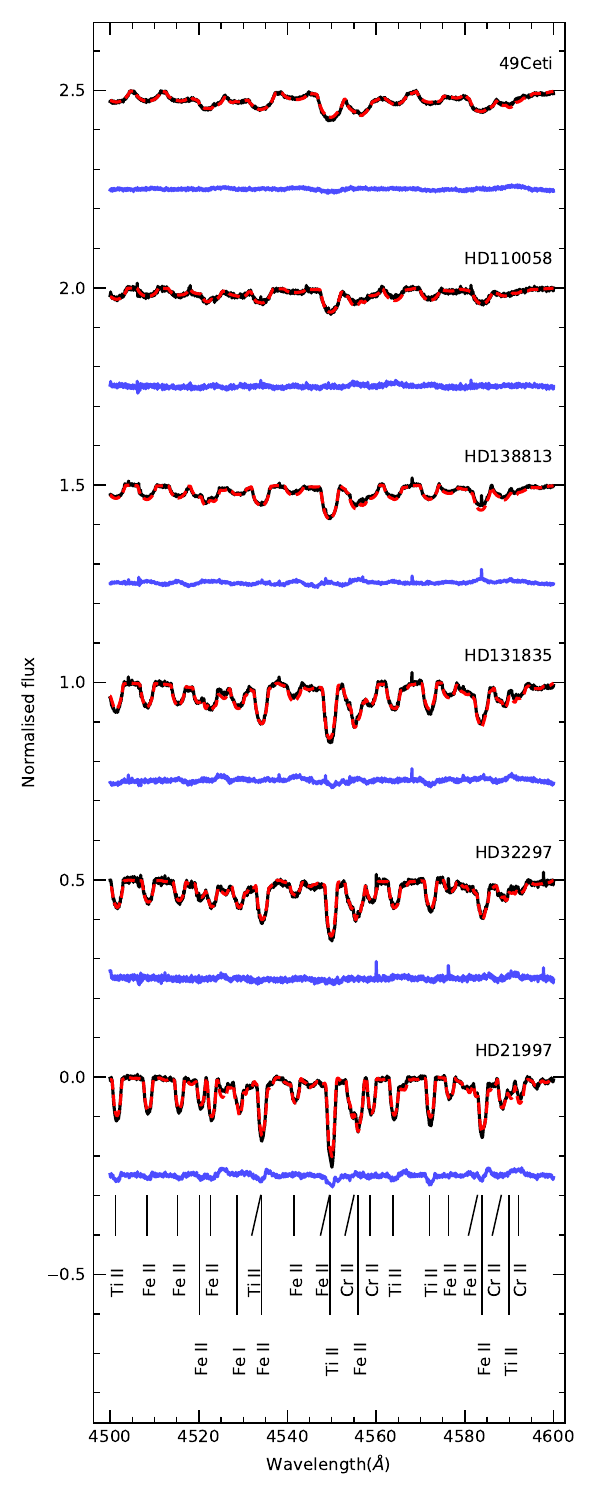}}
 \caption{Spectral fitting of the debris disk host star sample. The spectra are shifted vertically for better visualisation. Black represents the observed spectrum, red represents the best-fit synthetic spectrum, and blue represents the difference between the observed and synthetic spectrum. The spectra are sorted in increasing order of \vsini\ value from bottom to top.}
 \label{fig:spectral_fitting}
\end{figure}

\begin{figure*}
\includegraphics[width=17cm,trim=1.0cm 0.5cm 3.0cm 0.0cm]{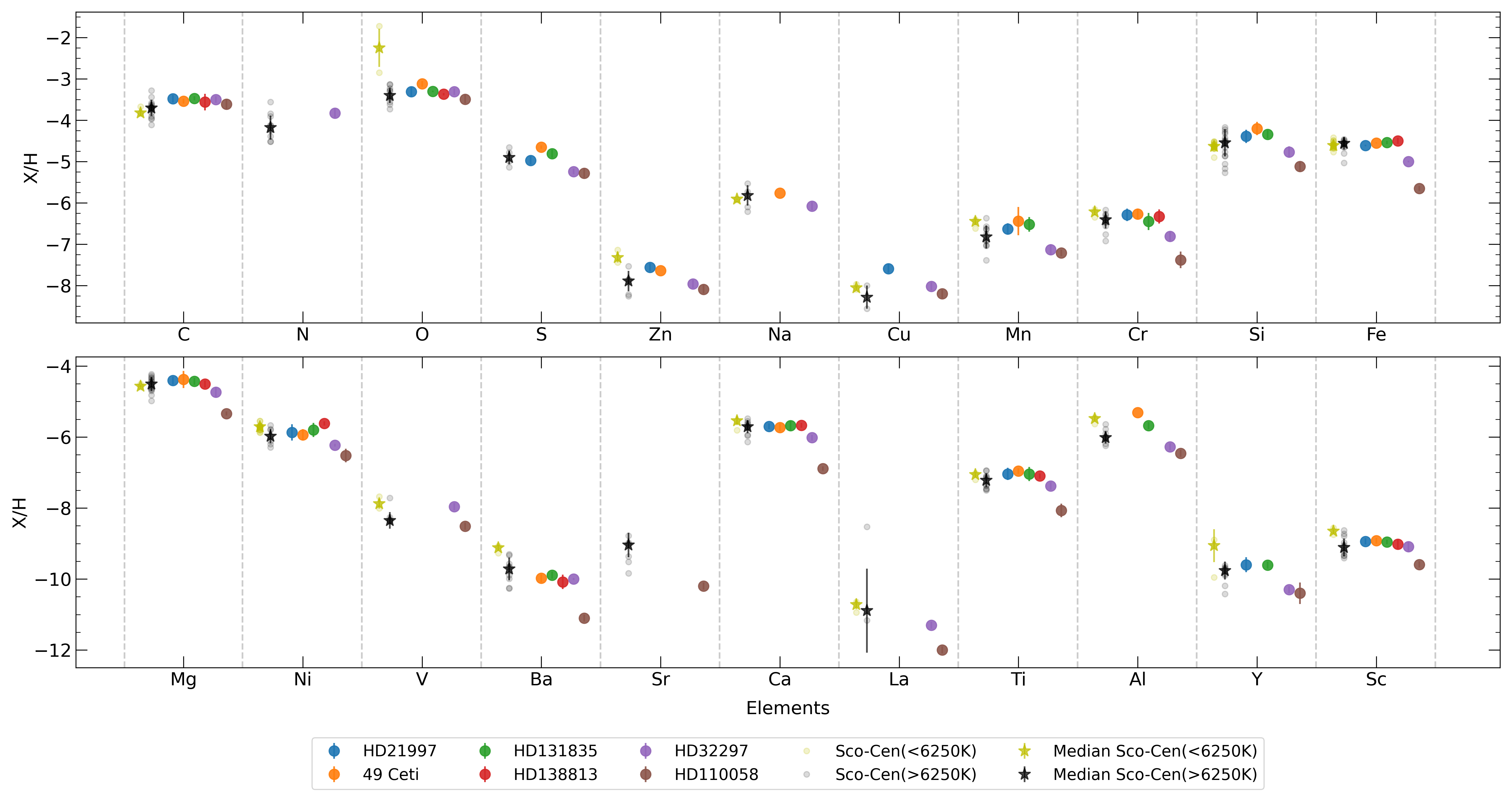}
 \caption{Comparing the elemental abundances of debris disk host stars (this work) with that of similar stars in the Sco-Cen association (see Table\,\ref{tab:sco-cen_stars}) as reference. The yellow and grey circles represent individual Sco-Cen association members. The stars represent the median abundance of the Sco-Cen association members. The $x$-axis is in increasing order of elemental condensation temperature. The condensation temperature also increases from the top to the bottom panel. The volatile abundances (C, N, O) show no depletion in all stars. The refractory abundances are depleted in \starfive\ and \starsix, likely due to dust-poor material accretion from their early protoplanetary disks.}

 \label{fig:comparison_abundnance}
\end{figure*}

In this section, we describe the stellar parameter and abundance estimates for the six stars in our sample. An example of our spectral fitting is shown in Fig.\,\ref{fig:spectral_fitting} and Fig. \ref{fig:spectral_fitting_C_and_O}, in a narrow wavelength range for the six stars. The resulting averaged best fit values are listed in Table\,\ref{tab:stellar_abun}.
The abundances are compared to the Sco-Cen reference abundance listed in Table\,\ref{tab:stellar_abun} and discussed in detail in Sect.\,\ref{sec:reference abundance}. The abundances for our sample of stars are plotted in Fig.\,\ref{fig:comparison_abundnance}, along with the Sco-Cen reference abundances (see Sec. \ref{sec:reference abundance}) for different elements.In the figure, the x-axis is ordered by increasing condensation temperature from left to right and from top to bottom.

\starone: This star has the lowest \vsini\ value in our sample. All abundances are close to our reference abundance from Sco-Cen association members.

\startwo: This star has the highest \vsini\ of $196.9\,$km/s. Most absorption lines are heavily blended, making the abundance estimates for some elements more uncertain. All the abundances are close to the Sco-Cen reference abundance,

\starthree: All elemental abundances are similar to the Sco-Cen reference abundance.

\starfour: This is the hottest star in our sample. The abundances are similar to the Sco-Cen reference abundance.

\starfive: It is the coolest star in our sample. The abundances show a high depletion of abundances of refractory elements. The volatile abundances are close to the Sco-Cen reference abundance. It is a $\lambda\,$Bo\"{o} type chemically peculiar star.

\starsix: This star also shows refractory depletion, whereas the volatile abundances are similar to the Sco-Cen reference abundance. It is a $\lambda\,$Bo\"{o} type chemically peculiar star.

\begin{table*}[t]
\renewcommand{\arraystretch}{1.3}
  \small
  \caption{Stellar parameters and elemental abundances for our sample of six debris disk-hosting stars. The median abundance of the hot Sco-Cen sample (Table\,\ref{tab:sco-cen_stars}) and the Solar abundances are also listed for reference.}
  \centering
  \begin{tabular}{c   c   c   c   c   c   c   c  c}
  \hline
  \hline
    Element & \starone & \startwo & \starthree & \starfour & \starfive & \starsix & Sco-Cen & Sun\\
    \hline
     C  & $-3.48\pm0.09$ & $-3.54\pm0.06$ & $-3.47\pm0.05$ & $-3.56\pm0.20^*$ & $-3.61\pm0.12$ & $-3.50\pm0.08$ & $-3.70 \pm 0.19$ & $-3.57$\\
     N & & & & & & $-3.83 \pm 0.20$ & $-4.17 \pm 0.29$ & $-4.17$\\
     O  & $-3.30\pm0.11$ & $-3.12\pm0.07$ & $-3.30\pm0.02$ & $-3.37\pm0.03$ & $-3.49\pm0.20^*$ & $-3.31\pm0.08$ & $-3.40 \pm 0.19$ & $-3.31$\\
    Na &    -    & $-5.77 \pm 0.20^*$ &    -    &    -    & - & $-6.08 \pm 0.20^*$ & $-5.82 \pm 0.25$ & $-5.76$\\
    Mg & $-4.41\pm0.11$ & $-4.39\pm0.24$ & $-4.43\pm0.10$ & $-4.51\pm0.10$ & $-5.34\pm0.06$ & $-4.74\pm0.14$ & $-4.51 \pm 0.20$ & $-4.40$\\
    Al &    -    & $-5.32 \pm 0.20^*$ &   $-5.68 \pm 0.20^*$     &    $-5.41 \pm 0.20^*$    & $-6.46 \pm 0.20^*$ & $-6.28 \pm 0.20^*$ & $-6.01 \pm 0.17$ & $-5.55$\\
    Si & $-4.39\pm0.16$ & $-4.20\pm0.16$ & $-4.34\pm0.10$ &     -   & $-5.12\pm0.11$ & $-4.77\pm0.14$ & $-4.54 \pm 0.33$ & $-4.49$\\
    S  & $-4.97\pm0.10$& $-4.65\pm0.04$ &$-4.81\pm0.01$&     -   & $-5.28\pm0.11$ & $-5.24\pm0.07$ & $-4.90 \pm 0.17$ & $-4.88$\\
    Ca & $-5.70\pm0.08$ & $-5.74\pm0.11$ & $-5.68\pm0.13$ & $-5.67\pm0.07$ & $-6.89\pm0.06$ & $-6.02\pm0.08$ & $-5.71 \pm 0.19$ & $-5.66$\\
    Sc & $-8.94\pm0.09$ & $-8.92\pm0.08$ & $-8.96\pm0.06$ &$-9.02\pm0.13$& $-9.59\pm0.09$ & $-9.09\pm0.14$ & $-9.11 \pm 0.25$ & $-8.85$\\
    Ti & $-7.03\pm0.17$ & $-6.97\pm0.08$ & $-7.04\pm0.20$ & $-7.10\pm0.07$ & $-8.07\pm0.19$ & $-7.38\pm0.11$ & $-7.22 \pm 0.19$ & $-7.05$\\
    V  &    -    &    -    &    -    &    -    & $-8.51 \pm 0.20^*$ & $-7.96 \pm 0.20^*$ & $-8.35 \pm 0.23$ & $-8.07$\\ 
    Cr & $-6.29\pm0.15$ & $-6.27\pm0.13$ & $-6.44\pm0.21$ & $-6.33\pm0.17$ & $-7.38\pm0.20$ & $-6.81\pm0.11$ & $-6.41 \pm 0.21$ & $-6.36$\\
    Mn & $-6.63\pm0.14$ & $-6.44\pm0.34$ & $-6.52\pm0.18$ &     -   & $-7.21\pm0.08$ & $-7.13\pm0.13$ & $-6.83 \pm 0.28$ & $-6.57$\\
    Fe & $-4.61\pm0.02$ & $-4.56\pm0.01$ & $-4.54\pm0.04$ & $-4.50\pm0.07$ & $-5.65\pm0.06$ & $-4.99\pm0.02$ & $-4.56 \pm 0.15$ & $-4.50$\\
    Ni & $-5.87\pm0.23$ & $-5.95\pm0.16$ & $-5.80\pm0.20$ & $-5.62\pm0.06$ & $-6.52\pm0.19$ & $-6.23\pm0.04$ & $-5.97 \pm 0.20$ & $-5.78$\\
    Cu & $-7.59 \pm 0.20^*$ &   -     &   -     &   -     & $-8.20 \pm 0.20^*$ & $-8.02 \pm 0.20^*$ & $-8.28 \pm 0.28$ & $-7.81$\\
    Zn & $-7.56 \pm 0.20^*$ & $-7.65 \pm 0.20^*$ &    -    &    -    & $-8.09 \pm 0.20^*$ & $-7.96 \pm 0.20^*$ & $-7.89 \pm 0.25$ & $-7.44$\\
    Sr &    -    &    -    &    -    &    -    &$-10.21 \pm 0.20^*$ & - & $-9.04 \pm 0.34$ & $-9.13$\\
    Y  & $-9.60\pm0.20$ &    -    & $-9.61 \pm 0.20^*$ &    -    & $-10.41\pm0.30$ & $-10.30 \pm 0.20^*$ & $-9.76 \pm 0.25$ & $-9.79$\\
    Ba &$-9.82\pm0.10$& $-9.98\pm0.11$ & $-9.88 \pm 0.20^*$ & $-10.08 \pm 0.20^*$& $-11.16 \pm 0.20^*$ & $-10.07\pm0.10$ & $-9.71 \pm 0.32$ & $-9.82$\\
    La &    -    &    -    &    -    &    -    &$-12.01 \pm 0.20^*$ & $-11.30 \pm 0.20^*$ & $-10.89 \pm 1.18$ & $-10.90$\\
    \hline
    \teff &  $8440\pm100$ &   $8970\pm100$    &  $8550\pm90$  &   $9300\pm90$    & $7880\pm80$ & $8060\pm90$ \\
    \logg &  $4.30\pm0.15$ &  $4.19\pm0.15$ &   $4.15\pm0.20$ & $4.22\pm0.17$ & $4.00\pm0.05$ & $4.25\pm0.13$ \\
    \vsini & $70.4\pm0.9$ & $196.9\pm2.1$ & $106.2\pm1.1$ & $136.2\pm1.5$ & $154.8\pm2.8$ & $90.1\pm0.4$ \\
    \vmic &  $2.9\pm0.4$ &  $2.5\pm0.2$ &  $3.0\pm0.5$ &  $2.4\pm0.3$ &  $3.0\pm0.2$ &  $2.9\pm0.2$ \\
    \hline\\
  \end{tabular}
  \tablefoot{Abundances are in $\log (N_{\rm el}/N_{\rm H})$. The $^*$ mark indicates abundance estimates from a single absorption line, with a typical uncertainty of 0.20 dex assigned. The Sco-Cen column is the median abundance of the hot star sample of Sco-Cen association member stars (right panel of Table\,\ref{tab:sco-cen_stars}). The uncertainty is based on the standard deviation of different stellar estimates.}
  \label{tab:stellar_abun}
\end{table*}

\begin{table}[t]
\renewcommand{\arraystretch}{1.3}
  \small
  \caption{Sco-Cen association members used from literature for reference abundance.}
  \centering
  \begin{tabular}{l r| l r}
    \hline
    \hline
    \multicolumn{1}{c}{Star} & \multicolumn{1}{c|}{Reference} & \multicolumn{1}{c}{Star} & \multicolumn{1}{c}{Reference} \\
    \hline
    \multicolumn{2}{c|}{\teff\, < 6250 K} & \multicolumn{2}{c}{\teff\, > 6250 K} \\
    \hline
    CP-52 5025 & 9 & HD 112810 & 1 \\
    CD-51 6900 & 9 & HD 135344 & 2 \\ 
    CD-36 10208 & 9 & HD 120326 & 3 \\
    CD-49 4947 & 9 & HD 133803 & 3 \\  
    CD-23 13281 & 9 & HD 144432 & 4 \\ 
    CD-39 10162 & 9 & HD 100453 & 2 \\ 
    CD-51 10295 & 9 & HD 142666 & 4 \\ 
    CD-22 11502 & 9 & HD 115820 & 3 \\ 
    TYC 9034 968 1 & 9 & HD 95086 & 3 \\ 
    CD-40 7581 & 9 & HD 139614 & 4 \\ 
    TYC7886 1894 1 & 9 & Eta Leo & 5 \\ 
    CD-25 11330 & 9 & HD 122980 & 6 \\ 
    CD-50 6815 & 10 & HR 4898 & 7 \\
    CD-34 10180 & 9 & HD 120324 & 8 \\ 
    CD-40 8031 & 9 & HR 5285 & 7 \\ 
    CD-49 7027 & 10 & HR 4853 & 7 \\ 
    CD-31 13486 & 9 & HR 6165 & 7 \\  
    HD 102458 & 1 & HD 149438 & 6 \\
    \hline
  \end{tabular}
  \tablefoot{Detailed stellar elemental abundances of each star are listed in Tables\,\ref{tab:Sco-Cen detailed abundance1}, \ref{tab:Sco-Cen detailed abundance2} and \ref{tab:Sco-Cen detailed abundance3}.}
  \tablebib{(1) \citet{2018BuderGALAHDR2}, (2) \citet{2004AckeWaelkens}, (3) \citet{2021Saffe}, (4) \citet{2012Folsom}, (5) \citet{2013Tanriverdi}, (6) \citet{2012Nievacosmicabun}, (7) \citet{1994Kilian}, (8) \citet{2017AhmedSigut}, (9) \citet{2009VianaSACY}, (10) \citet{2019SimpsonGALAH}.}
  \label{tab:sco-cen_stars}
\end{table}

The new stellar parameters and elemental abundances are analysed using the {\sc CAMstars}\footnote{\texttt{\href{https://github.com/adamjermyn/CAMstars}{https://github.com/adamjermyn/CAMstars}}} \citep{2018JermynCAMStars} accretion contamination model to derive constraints on the mass accretion history of individual systems. Investigating accretion contamination can reveal dust-poor accretion from a gas-rich debris disk or an earlier transitional protoplanetary disk with a strong dust trap. The accretion contamination from a protoplanetary disk with a strong dust trap would manifest as depletion of all elements that are sequestered away into trapped solids and thus unable to accrete with the viscous, volatile-rich gas disk.

\subsection{Contamination from accreting material}\label{subsec:fph description}
The contamination of the stellar photosphere by the accreting material is represented by the following equation
\begin{equation}
    \left(\frac{X}{H}\right)_{\rm obs} = f_{\rm ph}\left(\frac{X}{H}\right)_{\rm acc}+(1-f_{\rm ph})\left(\frac{X}{H}\right)_{\rm ref} ,
    \label{eqn:contamination equation}
\end{equation}\\
where
\begin{equation}
    f_\mathrm{ph} = f\left(\dot{M}_{\rm acc},{D_{\rm ph}}\right) .
\end{equation}
Here, \xhobs\ is the observed abundance measured in this paper (Table\,\ref{tab:stellar_abun}), \xhacc\ is the abundance of the accreting material, \xhref\ is the reference abundance representing the stellar abundance without accretion contamination, \fph\ is the accretion contamination parameter, \mdot\ is the gas accretion rate and $D_{\rm ph}$ is the diffusion coefficient. Although we report all the abundances in logarithmic terms, Eq. \ref{eqn:contamination equation} requires \xhobs, \xhacc\ and \xhref\ to be absolute ratios of number densities. We used {\sc CAMstars} to estimate \fph, which also provides the $D_{\rm ph}$ value for the star. The $D_{\rm ph}$ for an A-type star is dominated by rotational mixing, which is much slower than the convective mixing in cooler stars. Rotational mixing depends on stellar \teff, \logg\ and equatorial rotational velocity ($\nu$). Stars with higher $\nu$ will have more efficient mixing due to higher values of $D_{\rm ph}$, and thus, they will have a lower \fph. Since we could only estimate \vsini\ ($\leq \nu$), we can only get an upper limit on the \fph\ value (Table\,\ref{tab:stardiskdata}).

To calculate the \fph\ values of our stars, {\sc CAMstars} also needs their \mdot\,(Table\,\ref{tab:stardiskdata}). We used the model described in \citet{Kraletal2017} to calculate the gas accretion rates onto the stars. In this model, the gas in the debris disk is considered secondary (produced by planetesimal collisions) because any primordial gas would be photodissociated in a timescale of $\sim 100$ years from the interstellar UV radiation and will not be detectable \citep{Kraletal2017}. The gas is continuously released in the planetesimal belt, which can then spread onto the star like a typical accretion disk. In the steady state, the gas release rate from the planetesimal belt is the accretion rate. The accretion rate can then be represented as \mdot = $M_{\rm CO}/t_{\rm s}$ where $M_{\rm CO}$ is the gas mass observed for the disk and $t_{\rm s}$ is the survival timescale of the gas before it dissociates. We adjust the gas release rate to fit the observed gas mass for the disk. We estimated the CO gas mass by fitting the CO lines assuming LTE. We used the ALMA observations of $\rm ^{12}CO$ and if available also its isotopologues (e.g. $\rm ^{13}CO$ and $\rm C^{18}O$). In case neutral C observations are available from ALMA, we use that as well. For the most massive disks, we account for neutral carbon shielding that can enhance the CO lifetime to fit the gas release rate \citep{2019KralHD131835}. Hence, in the most massive disks ($\gtrsim 10^{-2}\,M_\oplus$), CO can be shielded sufficiently to start spreading onto the star together with carbon and oxygen. Instead, in less massive disks, only the atomic species are expected to be able to reach the star. We note that a fraction of the gas could also be accreted by planets located between the belt and the star \citep{2020Kralpla}, which is not accounted for here. 

To calculate the uncertainty on \mdot, we have to account for the uncertainty on $M_{\rm CO}$ and $t_{\rm s}$, where the later dominates. The survival timescale for the gas, $t_{\rm s}$ depends on the exact quantity of CO, carbon and their radial distributions, which can affect CO self-shielding or carbon-shielding, at least for the most massive disks. Refining these estimates requires high-resolution imaging, such as with ALMA. Although it is difficult to assign a formal uncertainty to the accretion rates, previous work by \citet{2019KralHD131835} on \starthree\ suggests a conservative uncertainty of a factor of 10. For more massive disks, the CO mass is usually derived from isotopologues, but accurately determining $\rm ^{12}CO$ mass requires knowledge of isotopologue ratios, which remain uncertain; we currently assume ISM ratios. Since \fph\ is directly proportional to \mdot\ \citep{2018JermynCAMStars}, its values will also have a factor of 10 uncertainty.

\subsection{Reference abundance}\label{sec:reference abundance}
To quantify the change in the stellar abundances due to accretion, we need a reference abundance indicating the stellar abundance without contamination. Since all stars in our sample are in the solar neighbourhood ($d\lesssim150\,$pc), we can use the typical composition of nearby stars as our \xhref. The abundance of He, C, N, O, Ne, Mg, Si and Fe is uniform in stars within the solar neighbourhood \citep{2012Nievacosmicabun}. Furthermore, three out of six stars in our sample belong to the Sco-Cen association ($d\approx130\,$pc). Based on these considerations, we used the composition of Sco-Cen association members from the literature (Table\,\ref{tab:sco-cen_stars}) to extend the list of reference abundances to other elements. We divided the sample into cool (\teff\,$<6250\,$K, convective envelope) and hot (\teff\,$> 6250\,$K, radiative envelope) stars. The results are independent of the exact location of this boundary since the hottest star in the cool sample is $5700\,$K and the coolest star in the hot sample is $6500\,$K. We used the median abundance of each element from the hot star sample in Sco-Cen (Table\,\ref{tab:sco-cen_stars}) as the \xhref\ for our sample. Taking the median abundance of each element prevents the effect of any outliers, which in our case is due to $\lambda\,$Bo\"{o} type peculiar stars (e.g. - HD\,139614 and HD\,142666, Table \ref{tab:Sco-Cen detailed abundance2}) in the Sco-Cen sample. Table\,\ref{tab:sco-cen_stars} also lists the references for the abundance values for both the hot and cool Sco-Cen sample. Figure\,\ref{fig:comparison_abundnance} compares the Sco-Cen abundances with the debris disk hosting stars analysed here. The uncertainty on the reference abundance is the standard deviation on the abundances of the Sco-Cen members (Table \ref{tab:sco-cen_stars}).

\section{Discussion}\label{sec:discussion}

\subsection{Composition vs age}
We compared the abundances of protoplanetary disk-hosting stars \citep[Table \ref{tab:Protoplanetary disk ref abundance}]{2012Folsom, 2015Kama} with the debris disk-hosting stars as a function of age.

\begin{figure}
 \resizebox{\hsize}{!}{\includegraphics[trim=0.3cm 0.0cm 0.1cm 0.0cm,clip]{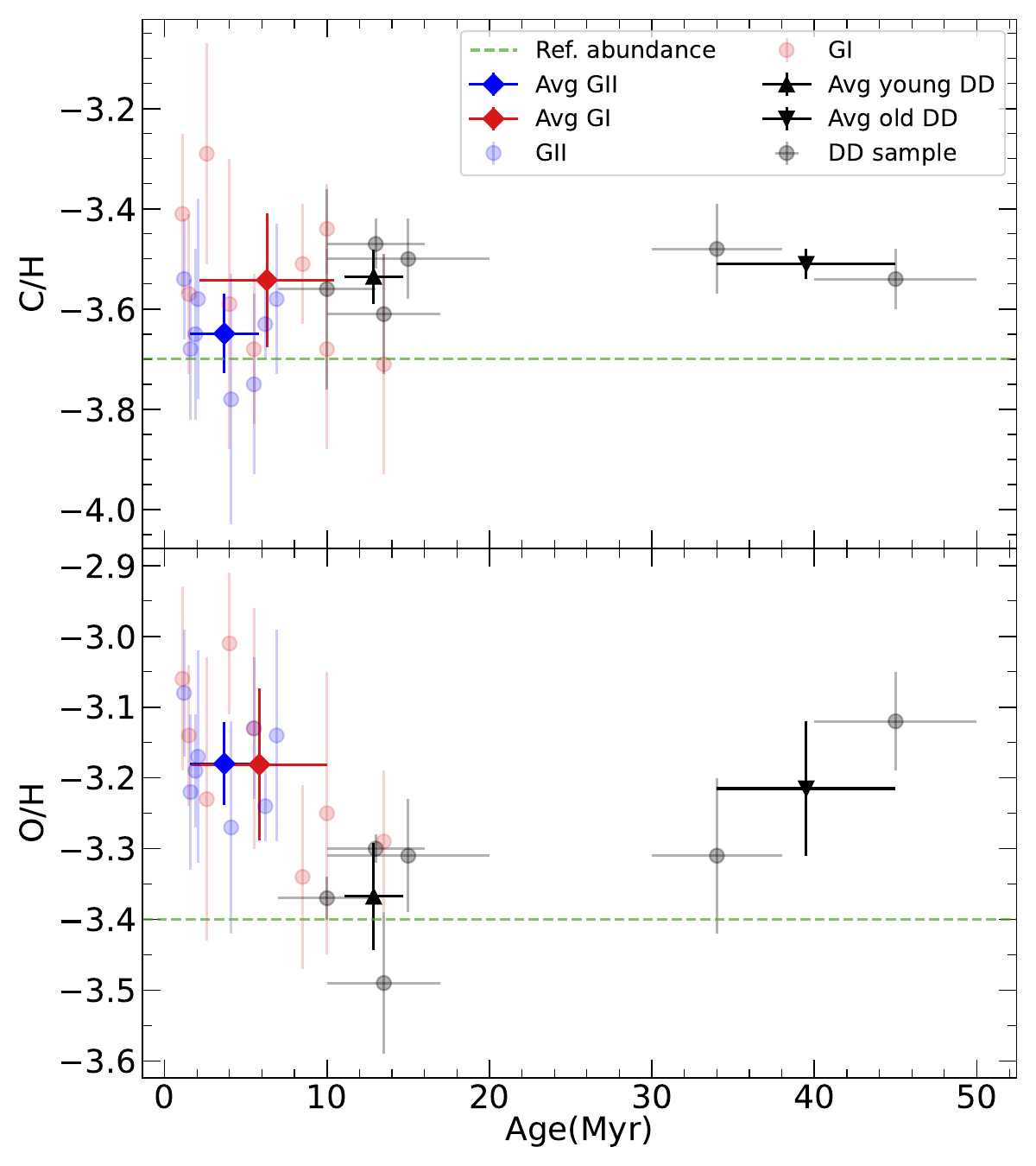}}
 \caption{Volatile abundance vs age plots. Top panel: C  vs age plot, Bottom panel: O vs age plot. GI and GII mean group\,I and group\,II protoplanetary disk-hosting stars, respectively. DD means debris disk-hosting stars. 
 }
 \label{fig:volatile_vs_age}
\end{figure}

Volatiles vs age: The volatile elements considered here are C and O. The C abundance vs age plot in the top panel of Fig.\,\ref{fig:volatile_vs_age} indicates that the average carbon abundance in stars hosting protoplanetary disks and debris disks is similar. The O abundances vs age plot in the bottom panel of Fig.\,\ref{fig:volatile_vs_age} reveals that young debris disk-hosting stars ($\rm <20\,Myr$) in our sample have a slightly lower average oxygen abundance compared to protoplanetary disk-hosting stars. The reason for the downward trend of the O abundance with age for the young stars is still an open question. The trend might be related to the accumulation of $\rm H_2O$ ice in dust traps, prohibiting the accretion of O. Alternatively, the formation of hydrated planetesimals (e.g. via streaming instability) beyond the snowline can reduce the amount of O available to accrete onto the star, thus depleting the stellar O abundance. These hydrated planetesimals (or asteroids) are observed in our Solar system and have been theorised to provide water to the solar system planets \citep{2024Kral}. The water can also accrete onto the star, potentially increasing O abundance during the debris disk stage \citep{2024Kral}. The potential upward trend after the young debris disk stage can be explained by the rotational mixing of the previously accreted material at the stellar photosphere with the inner layers of the envelope which will bring the observed O abundance to the actual stellar abundance without the contamination.

\begin{figure}
 \resizebox{\hsize}{!}{\includegraphics[trim=0.0cm 0.0cm 0.0cm 0.0cm,clip]{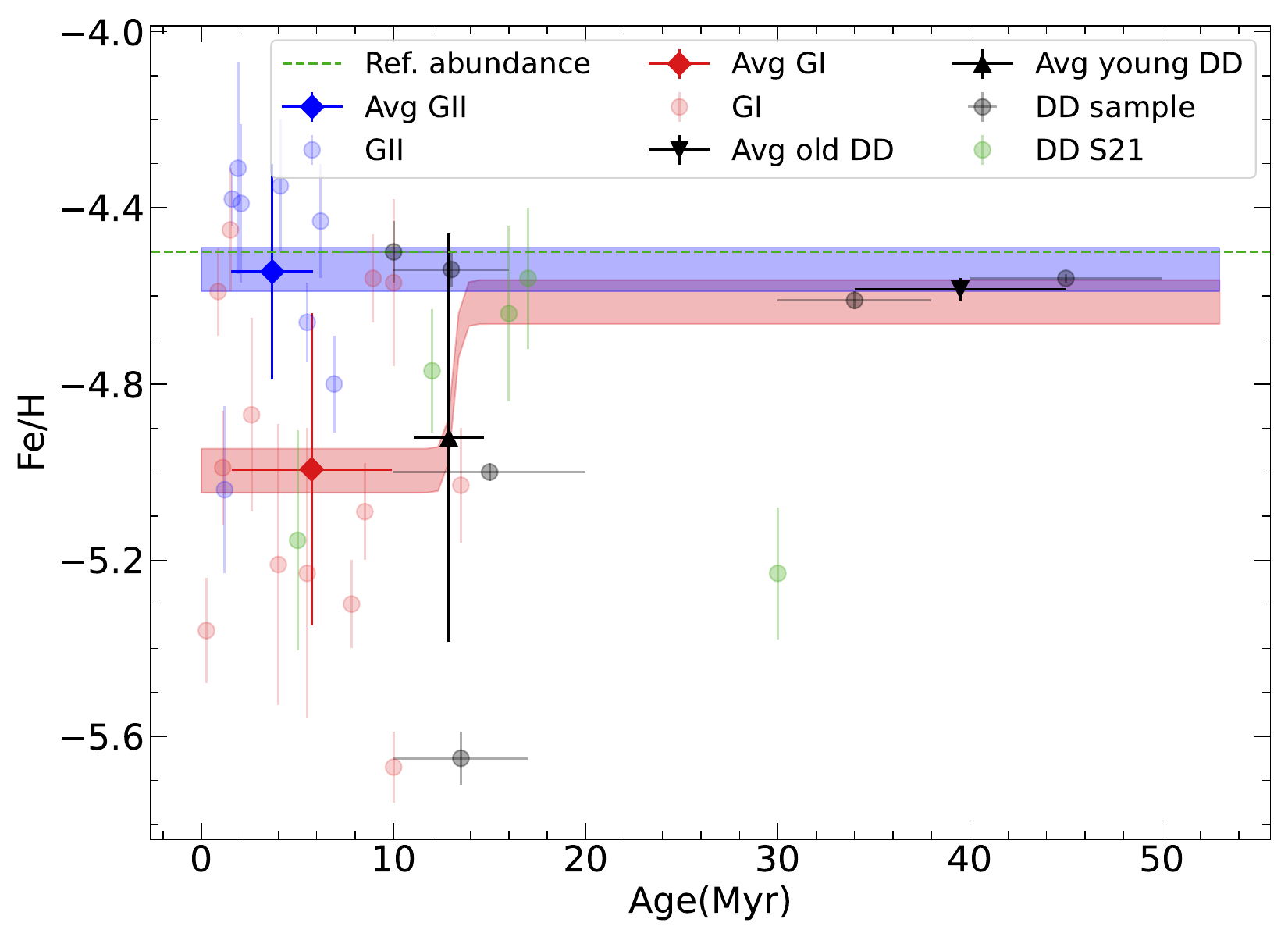}}
 \caption{Fe abundance vs age plot. Red curve and blue curve represents evolution of stellar Fe abundance for stars with group\,I and group\,II protoplanetary disks, respectively. DD S21: Young debris disk hosting stars HD\,169142, HD\,120326, $\rm \beta$ Pic, HR\,8799 and HD\,95086 from \citet{2021Saffe}. The abbreviations are described in Fig.\,\ref{fig:volatile_vs_age}.}
 \label{fig:Fe_vs_age}
\end{figure}

Refractories vs age: Fe is used as the representative for all the refractory elements in our sample. The Fe abundance vs age plot shown in Fig.\,\ref{fig:Fe_vs_age} indicates that the average elemental composition of the young debris disk-hosting stars is similar to that of the group\,I protoplanetary disk-hosting stars. Instead, the older stars in our sample have abundances close to our reference abundance (and the group II protoplanetary disk host stars) within the uncertainties. The young debris disk-hosting stars show a wide range of Fe abundances, from values similar to the reference abundance to being depleted by up to $\sim$ 1 dex. This is similar to group I protoplanetary disk-hosting stars, which are on average Fe-depleted, though some match the reference abundances.

\subsection{A possible depletion signature from the protoplanetary disk stage}
The theoretical \fph\ values calculated for all six stars are very low (last column in Table\,\ref{tab:stardiskdata}) due to the low accretion rates. This implies that the current debris disks cannot contaminate the stellar photospheres. Even if the \fph\ values increased by a factor of 10, equivalent to the typical uncertainty, the change in the observed abundance would be $\le0.01$ dex, which is below the accuracy of our abundance measurements. The depletion of refractory elements we observe for some of the stars in our sample, as shown in Fig.\,\ref{fig:comparison_abundnance}, needs an alternative explanation. The accretion rates for most Herbig\,Ae/Be stars are ${\sim}10^{-8}$ to ${\sim}10^{-6}\,$\msun\,/yr \citep{2023GrantAccRate}. Comparing that to the accretion rates of these debris disk hosting stars, which are in the order of $10^{-12}\,$\msun\,/yr, we propose that the depletion of refractory elements in our sample stars is a signature from the material that was accreted during their protoplanetary disk stage. Such a ``chronic depletion'' inherited from the gas-rich protoplanetary disk stage has been suggested before as an explanation for severely refractory-poor early-type stars, such as the $\lambda\,$Bo\"{o} chemically peculiar stars \citep{2015Kama, 2018JermynCAMStars}. 

The origin of this refractory depletion is associated with the accretion of dust-poor material. The formation of giant planets opens up gaps in the protoplanetary disks (group\,I disks). The gap produces a pressure maxima at its outer edge, preventing the heavier mm-sized dust grains from moving inward while allowing the gas to flow freely and accrete onto the star \citep{2006Rice,2012Pinilla}. This dust-poor accretion produces the refractory depletion in stellar abundances of the group\,I disk hosts. Stars hosting protoplanetary disks with no gaps (group\,II disks) have no change in their stellar abundances because they accrete both dust and gas. Figure\,\ref{fig:Fe_vs_age} shows the evolution of the stellar abundances for both group\,I (red) and group\,II (blue) protoplanetary disk hosting stars. Once the protoplanetary disk evolves into the debris disk, the accretion rate is too low to contaminate the stellar photosphere. Then stellar rotational mixing dominates and the refractory depleted abundances observed in group\,I disk hosts slowly disappear within $\sim 20\,$Myr \citep{2018JermynCAMStars}. The evolution curves in the figure illustrate this idea qualitatively, for the refractory stellar abundances in group I and group II protoplanetary disk host stars.

\subsection{Different accretion rates of \starfive}\label{subsec:accretion rate of hd110058}
\begin{figure}
 \resizebox{\hsize}{!}{\includegraphics[trim=0.0cm 0.1cm 0.1cm 0.2cm,clip]{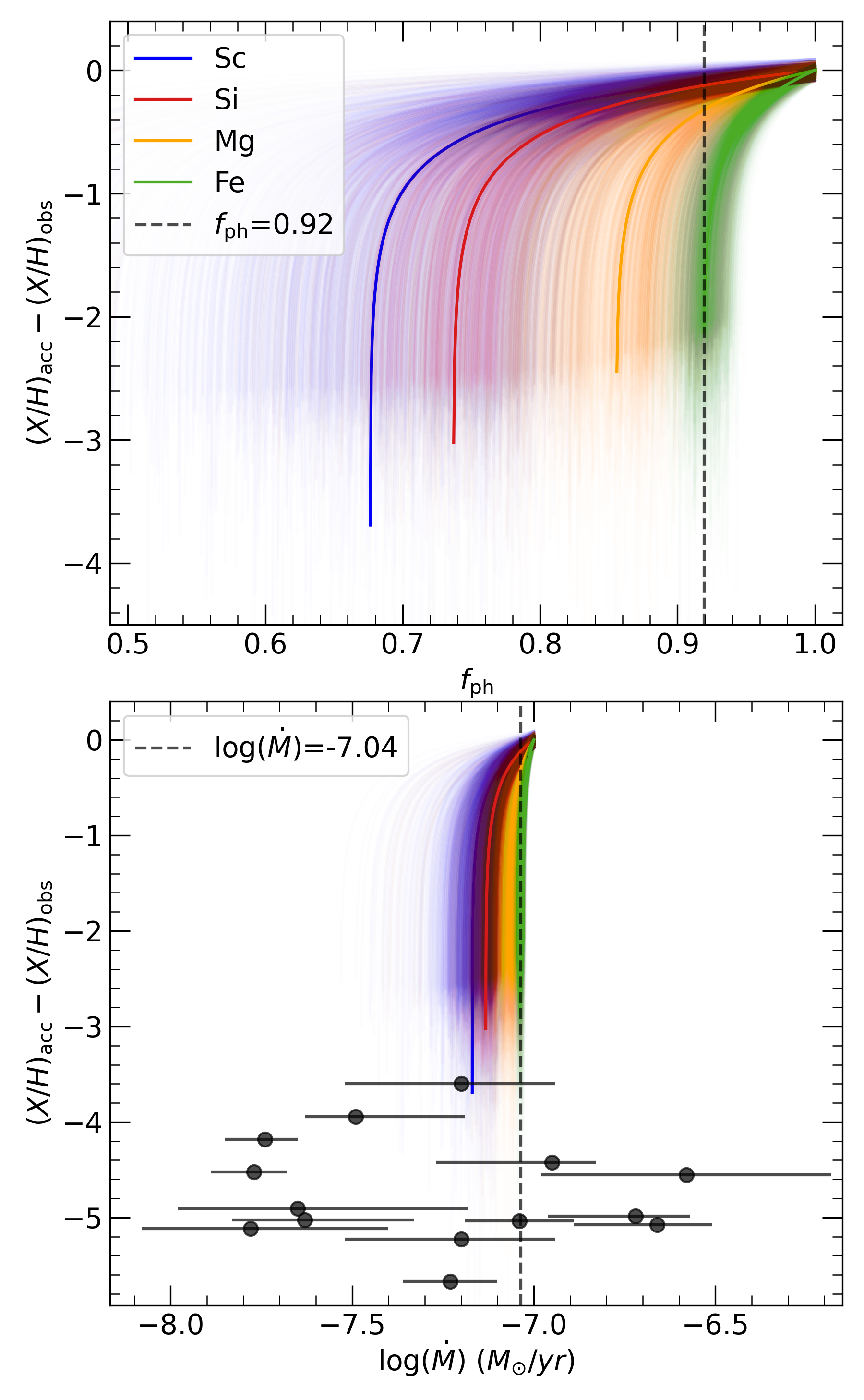}}
 \caption{ Range of \xhacc - \xhobs\ and \fph\ (top panel) or \mdot\ (bottom panel) satisfying Eq. \ref{eqn:fph range equation} for different elements. The y-axis for both panels is represented as \xhacc - \xhobs\ to bring all the elements to the same scale. The y-axis is logarithmic. The thin curves for each element represent range calculations considering uncertainty in \xhobs from our measurements and the dark curves represent the range for the average \xhobs. The dashed line represents the highest lower limit on the \fph\,(top panel) and \mdot\,(bottom panel). The black dots in the bottom panel represent \mdot\ and $\rm \left(\frac{Fe}{H}\right)_{acc}$ of Herbig stars from literature work summarised in \citet{2019Kama}.
 }
 \label{fig:lower_limit_accrate}
\end{figure}

\starfive\ shows the highest depletion in refractory elements in our sample. The theoretically calculated \fph\ for this star is almost zero ($1.42 \times 10^{-6}$, Table \ref{tab:stardiskdata}) due to the low accretion rate from the debris disk. This near-zero \fph\ value implies that the debris disk accretion cannot alter the bulk composition of the photosphere. On the other hand, substituting the observed and reference abundances in  Eq.\,\ref{eqn:contamination equation}, we can calculate the range of observed \fph\ and $\left(\frac{X}{H}\right)_{\rm acc}$ values that satisfy the equation 
\begin{equation}
    \left(\frac{X}{H}\right)_{\rm acc} = \frac{1}{f_{\rm ph}}\left[\left(\frac{X}{H}\right)_{\rm obs} - \left(\frac{X}{H}\right)_{\rm ref}\right] + \left(\frac{X}{H}\right)_{\rm ref}\,.
    \label{eqn:fph range equation}
\end{equation}
The top panel of Fig.\,\ref{fig:lower_limit_accrate} plots the range of different \xhacc and \fph\ values that satisfy Eq. \ref{eqn:fph range equation} for different refractory elements. The equation is satisfied for the range of \fph\ values, which give a positive \xhacc. Since the abundances used in the equation are in absolute number density ratios and not logarithms (Sect.\,\ref{subsec:fph description}), \xhacc\ in the equation cannot be negative. 

To interpret the results from the \fph\ range calculations from different elements, we will take the example of Sc and Fe. Sc gives a range of \fph\ values from $0.68 - 1$, which means at least 68\% of the stellar photosphere is contaminated by the accreting material. Whereas Fe gives us a much stronger constraint on the \fph\ value of $0.92 - 1$, implying at least 92\% of the photosphere is contaminated. Of all the elements, Fe gives us the highest lower limit on the observed \fph. This lower limit on the observed \fph\ value of 0.92 is quite different from the theoretical \fph\ value of $1.42 \times 10^{-6}$ (Table \ref{tab:stardiskdata}), calculated based on the current accretion rate and diffusion coefficient.

For each element, we also considered the uncertainty in our measured \xhobs\ while calculating the range of \xhacc\ and \fph\ values. This is shown as different curves for each element in Fig.\,\ref{fig:lower_limit_accrate}. As discussed in Sect.\,\ref{subsec:fph description}, \fph\ parameter depends on the equatorial rotational velocity of the star. Since we only measure \vsini, we need an estimate of the stellar inclination angle, $i$. The debris disk of \starfive\ is almost edge-on \citep{2022HalesALMAHD110058}. Assuming that the star-disk system is aligned, we can consider $i \approx 90^{\circ}$, implying our \vsini\ estimate is the equatorial rotational velocity of the star.

The \fph\ parameter is a function of \mdot, so we can invert it to get a lower limit on \mdot\ to be $\sim 9\times10^{-8}\,$\msun\,/yr (bottom panel of Fig.\,\ref{fig:lower_limit_accrate}). This high accretion rate is comparable to the accretion rates of Herbig stars (bottom panel of Fig.\,\ref{fig:lower_limit_accrate}). Comparing this value to the current accretion rate of \starfive\ of $1.42 \times 10^{-13}\,$\msun\,/yr implies that the refractory depleted material contaminating the stellar photosphere of \starfive\ is most likely to be from its protoplanetary disk stage.

\subsection{Comparison between \starcomp\ and \starfive}

The planetesimal belts which supply debris disks likely originate from massive dust rings in protoplanetary disks, where planetesimals can form in large numbers via the streaming instability. The instability produces pebble overdensities that can be unstable to direct gravitational collapse into large planetesimals \citep{2005Youdin_and_Goodman, 2009Johansen}. It is, therefore, interesting to note the following links between two similarly-aged, similar-type stars in Sco-Cen that have disks apparently on opposite sides of the protoplanetary vs debris disk boundary.

\begin{figure}
 \resizebox{\hsize}{!}{\includegraphics[trim=0.0cm 0.0cm 0.0cm 0.0cm,clip]{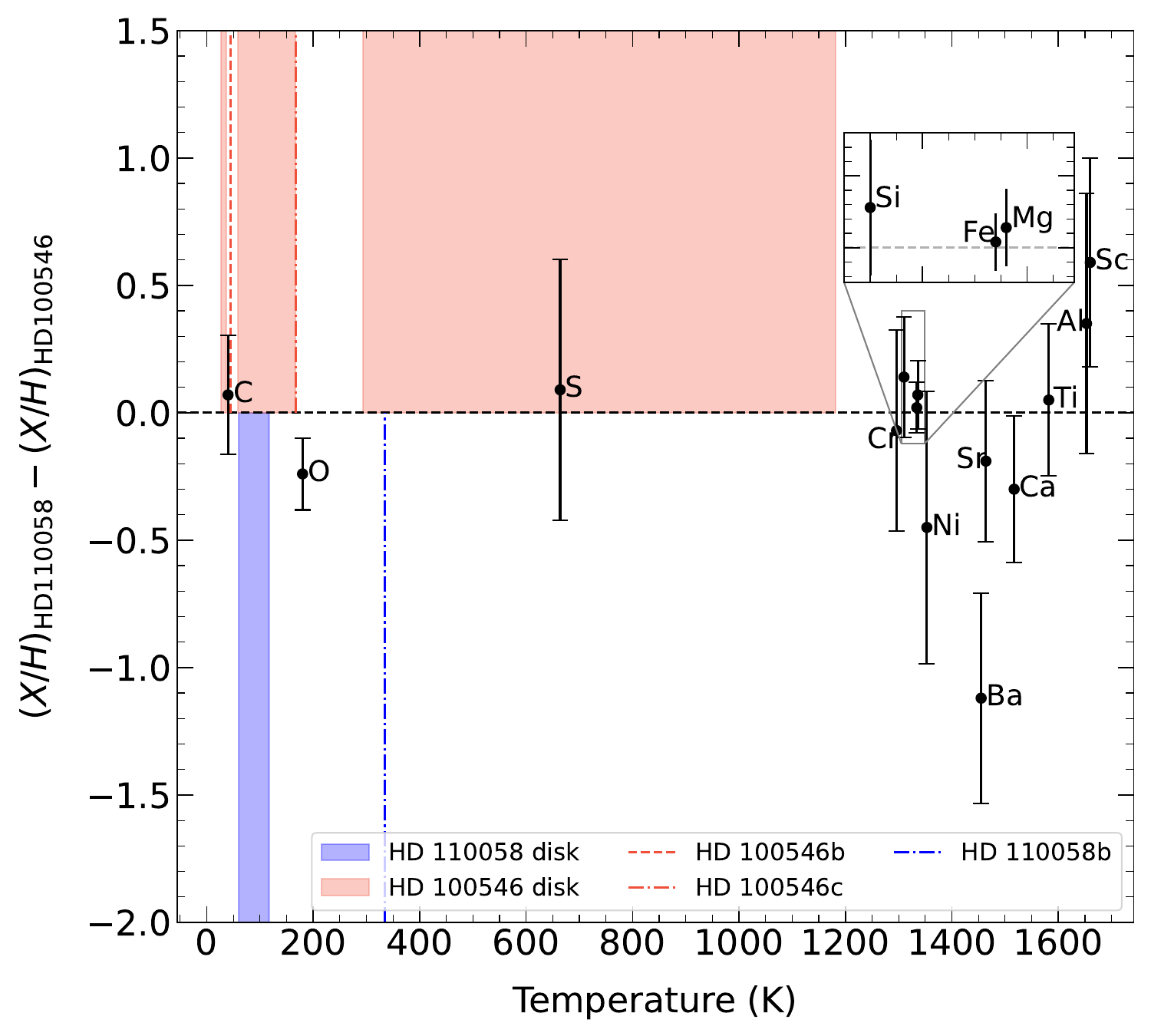}}
 \caption{Difference between stellar abundances of \starfive\ and \starcomp\ in increasing order of condensation temperature. The abundances are on a logarithmic scale. The highlighted regions in red (top) and blue (bottom) represent radial disk mid-plane dust temperatures for \starcomp\ and \starfive, respectively. For \starcomp, we used the mid-planet dust radial temperature profile from \citet{2024Keyte}. For \starfive, we used the Stefan Boltzmann law and stellar luminosity to estimate the radial temperature profile for the debris disk and giant planet. The vertical lines indicate the temperatures at the position of the planets (confirmed or predicted) in the radial disk temperature profiles around both stars.
 }
 \label{fig:diff_abun}
\end{figure}

\begin{figure}
 \resizebox{\hsize}{!}{\includegraphics[trim=0.0cm 0.0cm 0.0cm 4.0cm,clip]{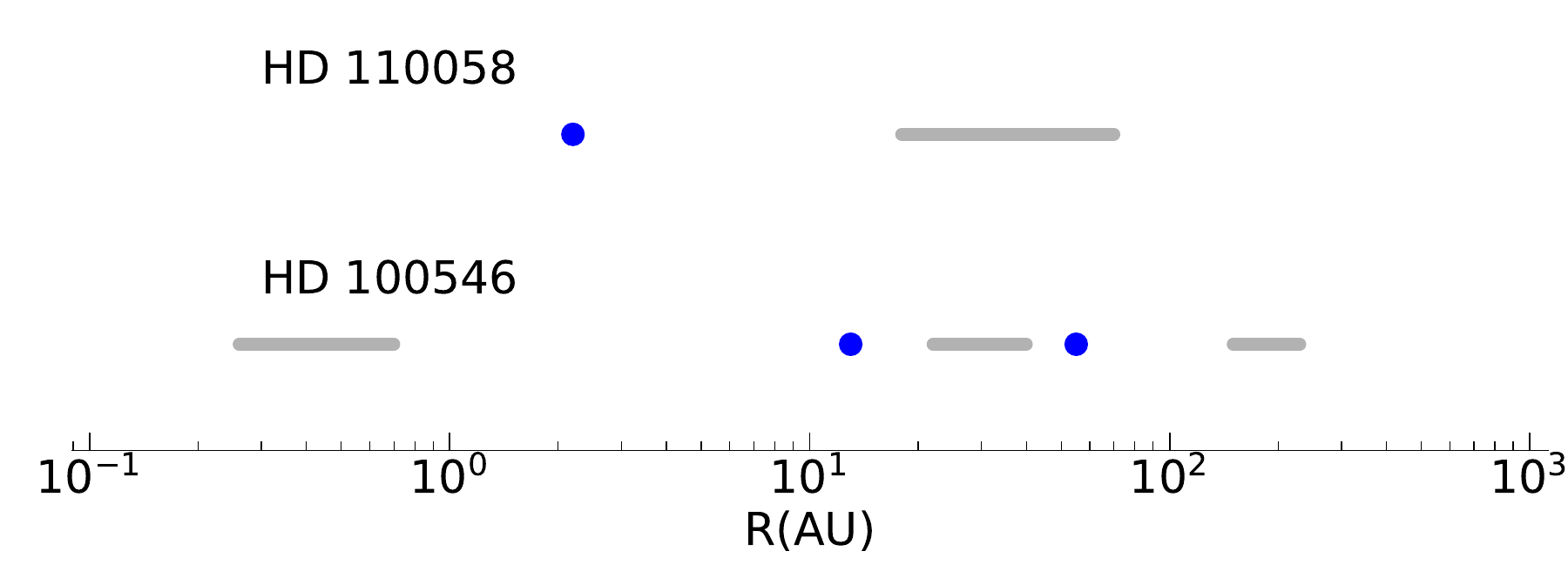}}
 \caption{Comparison of the dust disk positions (grey bars) between \starcomp\ and \starfive. The x-axis is in logarithmic scale. Blue dots represent confirmed and predicted giant planets.}
 \label{fig:compare_pp_and_dd_stars}
\end{figure}

\starcomp\ is a protoplanetary disk hosting star with an effective stellar temperature of \teff\,$=10\,500\,$K \citep{2016Kama}. It belongs to the Lower Centaurus Crux (LCC) of the Sco-Cen OB association \citep{2015Jang-CondellScoCenMember}. The main features of the dust disk are an inner cavity with a radius of ${\approx}\,22\,$au and two main dust rings, from $22$ to $40\,$au and $150$ to $230\,$au \citep{2024Keyte}. There is another small dust ring at $\lesssim1\,$au \citep{2024Keyte}, which is shown in Fig. \ref{fig:diff_abun} and Fig. \ref{fig:compare_pp_and_dd_stars}, but is not considered in this discussion. The refractory stellar abundances of \starcomp\ are the lowest in the protoplanetary disk hosting star sample assembled from \citet{2012Folsom} and \citet{2015Kama}, and the abundances are comparable to those of \starfive\ (Fig.\,\ref{fig:diff_abun}).

Since both \starcomp\ and \starfive\ belong to LCC, the composition of the initial material from which both these systems were formed should be similar. The calculated accretion rate of \starcomp\ is $9.12_{-1.37}^{1.18}\times10^{-8}\,$\msun\,/yr \citep{2015Fairlamb}, which is similar to the earlier protoplanetary disk accretion rate of \starfive\ (see Sect.\,\ref{subsec:accretion rate of hd110058}) within the uncertainties. Both stars are also suspected to have a Jupiter-sized planet in the innermost regions of the disk
\citep[Fig.\,\ref{fig:compare_pp_and_dd_stars}]{2015Pinilla, Stasevic2023HD110058}.  These similarities might indicate that the inner disk around \starfive\ is the next stage of evolution of \starcomp\ and thus, \starfive\ has recently transitioned into the debris disk phase. The similar depletion in refractory abundances (Fig.\,\ref{fig:diff_abun}) in both stars can be due to similar strengths in dust traps at similar locations (Fig.\,\ref{fig:compare_pp_and_dd_stars}). The strength and location of the dust trap in both stars might be associated with the previously predicted innermost giant planet.

The innermost dust ring of \starcomp\ will likely evolve into a planetesimal belt, and then potentially into a debris disk ring similar to that of \starfive. If the dust trap strength was similar in both stars, the dust mass available in the dust ring of \starfive\ to form planetesimals might also be similar. The dust mass at the innermost dust ring of \starcomp\ is $\sim 241.7\, M_{\oplus}$ \citep{2024Keyte}. There are two different dust mass estimates for the debris disk around \starfive. Below, we discuss the possible interpretations considering each dust mass estimate.

The secondary dust mass estimate for \starfive\ by \citet{2016SifryHD138813HD110058ALMA} is $0.003\,M_{\oplus}$. Assuming collisional equilibrium in the debris disk, a maximum size of 100 km for the planetesimals, and a standard Dohnanyi size distribution \citep{1969Dohnanyi} exponent of 3.5, we can indirectly infer the total planetesimal mass to be $\sim 30\,M_{\oplus}$. Comparing the dust mass in the innermost ring of \starcomp\ with the planetesimal mass in the debris disk of \starfive, we can calculate the efficiency of planetesimal formation, $\epsilon \approx 12 \%$. \citet{2009Johansen} simulated the formation of planetesimals in a protoplanetary disk through streaming instability and self-gravity and found $\epsilon = 20 \%$ with planetesimals of $100-200\,$km in size after five orbits. Further accretion becomes slow, thus slowing the growth of planetesimals. Although our results are an approximate estimate, it is interesting to point out the similarity in our results for the planetesimal formation efficiency ($\epsilon \approx12\,$\% from this observational work, $\epsilon =20\,$\% from the \citealt{2009Johansen} simulations). 

A more recent dust mass estimate of \starfive\ by \citet{2022HalesALMAHD110058} of $0.080^{+0.002}_{-0.003}\,M_{\oplus}$ is an order of magnitude higher than $0.003\,M_{\oplus}$. The inferred planetesimal mass in \starfive\ will then be $\sim 800\,M_{\oplus}$, which is higher than our considered initial dust mass for planetesimal formation ($\sim 241.7\, M_{\oplus}$, based on dust mass of inner dust ring of \starcomp). This might indicate that at least $800 - 241.7 \approx 558\,M_{\oplus}$ is already in planetesimals in the inner dust ring of \starcomp. 

\section{Conclusion} \label{sec:Conclusion}
We estimated the stellar parameters and elemental abundances of six debris disks hosting A-type stars using high-resolution and high S/N spectra. The main goals were to identify any chemical peculiarities, or lack thereof, and investigate any links between disk accretion and the stellar composition. Below, we summarise our key results and conclusions:

\begin{itemize}
    \item The oxygen abundance in young stars ($\rm <20\,Myr$) decreases with age until the debris disk stage, then may begin to rise again (Fig.\,\ref{fig:volatile_vs_age}). Whether the downward trend of O abundance is related to the accumulation of H$_{2}$O ice in dust traps or the formation of hydrated planetesimals via streaming instability, resulting in oxygen-poor accretion, remains an open question. The upward trend after the young debris disk stage is then related to rotational mixing at the stellar photosphere.
    
    \item \starone, \startwo, \starthree\ and  \starfour\ are chemically normal, while \starfive\ and \starsix\ show $\lambda\,$Bo\"{o}-type peculiarities due to refractory depletion in their photospheres.
    
    \item The low photospheric accretion contamination parameter (\fph) across all stars in our sample suggests ongoing debris disk accretion cannot significantly contaminate their photosphere. 
    
    \item The chronic depletion of refractory elemental abundances in \starfive\ and \starsix\ may indicate earlier contamination by dust-poor accretion from a protoplanetary disk. These stars may be the ``missing link'' between Herbig\,Ae/Be stars with dust-poor accretion and main-sequence $\lambda\,$Bo\"{o}-type field stars.

    \item We estimate a lower limit on the past accretion rate of $9 \times 10^{-8}\,M_{\odot}\,$/yr for \starfive, which is needed to explain the refractory depleted abundances observed in this star.
    This rate is similar to other Herbig stars but is five orders of magnitude higher than the current rate from the debris disk. This supports our hypothesis of previous accretion contamination of the stellar photosphere.

    \item \starfive\ shows refractory depletion similar to the protoplanetary disk-hosting star \starcomp, indicating comparable dust trapping in their disk. Both stars, part of the same Sco-Cen sub-association, likely formed from the same material. \starcomp\ has an accretion rate similar to the earlier accretion of \starfive. The depletion levels and solid belt locations suggest \starfive\ may be a more evolved version of \starcomp. 
    
\end{itemize}
Extending our abundance analysis to a larger sample of young debris disk-hosting stars will strengthen our hypothesis of the evolution of refractory stellar abundances with age due to stellar mixing. This will also offer insights into the timescale of the mixing process. Additionally, further investigation is needed to confirm the observed downward trend in oxygen abundance for young stars with disks and to identify its origin.
\color{black}

\begin{acknowledgements} 
  AA, SB, CPF, LF, and MK gratefully acknowledge funding from the European Union's Horizon Europe research and innovation programme under grant agreement No. 101079231 (EXOHOST) and from UK Research and Innovation (UKRI) under the UK government’s Horizon Europe funding guarantee (grant number 10051045). AA acknowledges support from the Estonian Research Council grant PRG 2159. This research has used the VizieR catalogue access tool, CDS, Strasbourg, France. We thank the anonymous referee whose comments improved the quality of our paper. We thank Denis Shulyak for providing access to {\sc LLmodels}. We thank Luke Kyte for providing the radial disk mid-plane temperature profile for HD 100546. We thank Heleri Ramler and Alexandra Lehmets for their useful comments on the paper draft.
\end{acknowledgements}

\bibliographystyle{aa} 
\bibliography{ref}      

\begin{thebibliography}{93}
\expandafter\ifx\csname natexlab\endcsname\relax\def\natexlab#1{#1}\fi

\bibitem[{{Acke} \& {Waelkens}(2004)}]{2004AckeWaelkens}
{Acke}, B. \& {Waelkens}, C. 2004, \aap, 427, 1009

\bibitem[{{Ahmed} \& {Sigut}(2017)}]{2017AhmedSigut}
{Ahmed}, A. \& {Sigut}, T.~A.~A. 2017, \mnras, 471, 3398

\bibitem[{{Anders} {et~al.}(2022){Anders}, {Khalatyan}, {Queiroz}, {Chiappini}, {Ard{\`e}vol}, {Casamiquela}, {Figueras}, {Jim{\'e}nez-Arranz}, {Jordi}, {Mongui{\'o}}, {Romero-G{\'o}mez}, {Altamirano}, {Antoja}, {Assaad}, {Cantat-Gaudin}, {Castro-Ginard}, {Enke}, {Girardi}, {Guiglion}, {Khan}, {Luri}, {Miglio}, {Minchev}, {Ramos}, {Santiago}, \& {Steinmetz}}]{2022AndersHD21997}
{Anders}, F., {Khalatyan}, A., {Queiroz}, A.~B.~A., {et~al.} 2022, \aap, 658, A91

\bibitem[{{Balona}(1994)}]{1994Balona}
{Balona}, L.~A. 1994, \mnras, 268, 119

\bibitem[{{Bessell} {et~al.}(1998){Bessell}, {Castelli}, \& {Plez}}]{1998Johnson}
{Bessell}, M.~S., {Castelli}, F., \& {Plez}, B. 1998, \aap, 333, 231

\bibitem[{{Buder} {et~al.}(2018){Buder}, {Asplund}, {Duong}, {Kos}, {Lind}, {Ness}, {Sharma}, {Bland-Hawthorn}, {Casey}, {de Silva}, {D'Orazi}, {Freeman}, {Lewis}, {Lin}, {Martell}, {Schlesinger}, {Simpson}, {Zucker}, {Zwitter}, {Amarsi}, {Anguiano}, {Carollo}, {Casagrande}, {{\v{C}}otar}, {Cottrell}, {da Costa}, {Gao}, {Hayden}, {Horner}, {Ireland}, {Kafle}, {Munari}, {Nataf}, {Nordlander}, {Stello}, {Ting}, {Traven}, {Watson}, {Wittenmyer}, {Wyse}, {Yong}, {Zinn}, {{\v{Z}}erjal}, \& {Galah Collaboration}}]{2018BuderGALAHDR2}
{Buder}, S., {Asplund}, M., {Duong}, L., {et~al.} 2018, \mnras, 478, 4513

\bibitem[{{Castelli} {et~al.}(1997){Castelli}, {Gratton}, \& {Kurucz}}]{1997Castelli}
{Castelli}, F., {Gratton}, R.~G., \& {Kurucz}, R.~L. 1997, \aap, 318, 841

\bibitem[{Castelli \& Kurucz(2004)}]{2004castelliKurucz}
Castelli, F. \& Kurucz, R.~L. 2004, New Grids of ATLAS9 Model Atmospheres

\bibitem[{{Cataldi} {et~al.}(2023){Cataldi}, {Aikawa}, {Iwasaki}, {Marino}, {Brandeker}, {Hales}, {Henning}, {Higuchi}, {Hughes}, {Janson}, {Kral}, {Matr{\`a}}, {Mo{\'o}r}, {Olofsson}, {Redfield}, \& {Roberge}}]{2023Cataldi}
{Cataldi}, G., {Aikawa}, Y., {Iwasaki}, K., {et~al.} 2023, \apj, 951, 111

\bibitem[{{Cataldi} {et~al.}(2020){Cataldi}, {Wu}, {Brandeker}, {Ohashi}, {Mo{\'o}r}, {Olofsson}, {{\'A}brah{\'a}m}, {Asensio-Torres}, {Cavallius}, {Dent}, {Grady}, {Henning}, {Higuchi}, {Hughes}, {Janson}, {Kamp}, {K{\'o}sp{\'a}l}, {Redfield}, {Roberge}, {Weinberger}, \& {Welsh}}]{2020CataldiHD32297}
{Cataldi}, G., {Wu}, Y., {Brandeker}, A., {et~al.} 2020, \apj, 892, 99

\bibitem[{{Dohnanyi}(1969)}]{1969Dohnanyi}
{Dohnanyi}, J.~S. 1969, \jgr, 74, 2531

\bibitem[{{Draper} {et~al.}(2018){Draper}, {Matthews}, {Venn}, {Lambert}, {Kennedy}, \& {Sitnova}}]{2018Draper}
{Draper}, Z.~H., {Matthews}, B., {Venn}, K., {et~al.} 2018, \apj, 857, 93

\bibitem[{{Draper} {et~al.}(2016){Draper}, {Matthews}, {Kennedy}, {Wyatt}, {Venn}, \& {Sibthorpe}}]{2016Draper}
{Draper}, Z.~H., {Matthews}, B.~C., {Kennedy}, G.~M., {et~al.} 2016, \mnras, 456, 459

\bibitem[{{Fairlamb} {et~al.}(2015){Fairlamb}, {Oudmaijer}, {Mendigut{\'\i}a}, {Ilee}, \& {van den Ancker}}]{2015Fairlamb}
{Fairlamb}, J.~R., {Oudmaijer}, R.~D., {Mendigut{\'\i}a}, I., {Ilee}, J.~D., \& {van den Ancker}, M.~E. 2015, \mnras, 453, 976

\bibitem[{{Feldt} {et~al.}(2017){Feldt}, {Olofsson}, {Boccaletti}, {Maire}, {Milli}, {Vigan}, {Langlois}, {Henning}, {Moor}, {Bonnefoy}, {Wahhaj}, {Desidera}, {Gratton}, {K{\'o}sp{\'a}l}, {Abraham}, {Menard}, {Chauvin}, {Lagrange}, {Mesa}, {Salter}, {Buenzli}, {Lannier}, {Perrot}, {Peretti}, \& {Sissa}}]{2017FeldtHD131835}
{Feldt}, M., {Olofsson}, J., {Boccaletti}, A., {et~al.} 2017, \aap, 601, A7

\bibitem[{{Folsom} {et~al.}(2012){Folsom}, {Bagnulo}, {Wade}, {Alecian}, {Landstreet}, {Marsden}, \& {Waite}}]{2012Folsom}
{Folsom}, C.~P., {Bagnulo}, S., {Wade}, G.~A., {et~al.} 2012, \mnras, 422, 2072

\bibitem[{{Galicher} {et~al.}(2016){Galicher}, {Marois}, {Macintosh}, {Zuckerman}, {Barman}, {Konopacky}, {Song}, {Patience}, {Lafreni{\`e}re}, {Doyon}, \& {Nielsen}}]{2016Galicher_age}
{Galicher}, R., {Marois}, C., {Macintosh}, B., {et~al.} 2016, \aap, 594, A63

\bibitem[{{Grant} {et~al.}(2023){Grant}, {Stapper}, {Hogerheijde}, {van Dishoeck}, {Brittain}, \& {Vioque}}]{2023GrantAccRate}
{Grant}, S.~L., {Stapper}, L.~M., {Hogerheijde}, M.~R., {et~al.} 2023, \aj, 166, 147

\bibitem[{{Gray}(1988)}]{Gray1988}
{Gray}, R.~O. 1988, \aj, 95, 220

\bibitem[{{Guzm{\'a}n-D{\'\i}az} {et~al.}(2023){Guzm{\'a}n-D{\'\i}az}, {Montesinos}, {Mendigut{\'\i}a}, {Kama}, {Meeus}, {Vioque}, {Oudmaijer}, \& {Villaver}}]{2023GuzmanDiaz}
{Guzm{\'a}n-D{\'\i}az}, J., {Montesinos}, B., {Mendigut{\'\i}a}, I., {et~al.} 2023, \aap, 671, A140

\bibitem[{{Hales} {et~al.}(2019){Hales}, {Gorti}, {Carpenter}, {Hughes}, \& {Flaherty}}]{2019HalesHD131835_HD138813}
{Hales}, A.~S., {Gorti}, U., {Carpenter}, J.~M., {Hughes}, M., \& {Flaherty}, K. 2019, \apj, 878, 113

\bibitem[{{Hales} {et~al.}(2022){Hales}, {Marino}, {Sheehan}, {Ulloa}, {P{\'e}rez}, {Matr{\`a}}, {Kral}, {Wyatt}, {Dent}, \& {Carpenter}}]{2022HalesALMAHD110058}
{Hales}, A.~S., {Marino}, S., {Sheehan}, P.~D., {et~al.} 2022, \apj, 940, 161

\bibitem[{{Herbig}(1960)}]{Herbig1960}
{Herbig}, G.~H. 1960, \apjs, 4, 337

\bibitem[{{Higuchi} {et~al.}(2020){Higuchi}, {K{\'o}sp{\'a}l}, {Mo{\'o}r}, {Nomura}, \& {Yamamoto}}]{2020HiguchiHD21997_49Ceti}
{Higuchi}, A.~E., {K{\'o}sp{\'a}l}, {\'A}., {Mo{\'o}r}, A., {Nomura}, H., \& {Yamamoto}, S. 2020, \apj, 905, 122

\bibitem[{{Higuchi} {et~al.}(2019){Higuchi}, {Saigo}, {Kobayashi}, {Iwasaki}, {Momose}, {Soon}, {Sakai}, {Kunitomo}, {Ishihara}, \& {Yamamoto}}]{2019Higuchi49Ceti}
{Higuchi}, A.~E., {Saigo}, K., {Kobayashi}, H., {et~al.} 2019, \apj, 883, 180

\bibitem[{{Higuchi} {et~al.}(2017){Higuchi}, {Sato}, {Tsukagoshi}, {Sakai}, {Iwasaki}, {Momose}, {Kobayashi}, {Ishihara}, {Watanabe}, {Kaneda}, \& {Yamamoto}}]{2017Higuchi49Ceti}
{Higuchi}, A.~E., {Sato}, A., {Tsukagoshi}, T., {et~al.} 2017, \apjl, 839, L14

\bibitem[{{Hughes} {et~al.}(2017){Hughes}, {Lieman-Sifry}, {Flaherty}, {Daley}, {Roberge}, {K{\'o}sp{\'a}l}, {Mo{\'o}r}, {Kamp}, {Wilner}, {Andrews}, {Kastner}, \& {{\'A}brah{\'a}m}}]{2017Hughes49Ceti}
{Hughes}, A.~M., {Lieman-Sifry}, J., {Flaherty}, K.~M., {et~al.} 2017, \apj, 839, 86

\bibitem[{{Jang-Condell} {et~al.}(2015){Jang-Condell}, {Chen}, {Mittal}, {Manoj}, {Watson}, {Lisse}, {Nesvold}, \& {Kuchner}}]{2015Jang-CondellScoCenMember}
{Jang-Condell}, H., {Chen}, C.~H., {Mittal}, T., {et~al.} 2015, \apj, 808, 167

\bibitem[{{Jermyn} \& {Kama}(2018)}]{2018JermynCAMStars}
{Jermyn}, A.~S. \& {Kama}, M. 2018, \mnras, 476, 4418

\bibitem[{{Johansen} {et~al.}(2009){Johansen}, {Youdin}, \& {Mac Low}}]{2009Johansen}
{Johansen}, A., {Youdin}, A., \& {Mac Low}, M.-M. 2009, \apjl, 704, L75

\bibitem[{{Kaiser}(2006)}]{2006Kaiser}
{Kaiser}, A. 2006, in Astronomical Society of the Pacific Conference Series, Vol. 349, Astrophysics of Variable Stars, ed. C.~{Aerts} \& C.~{Sterken}, 257

\bibitem[{{Kama} {et~al.}(2016){Kama}, {Bruderer}, {van Dishoeck}, {Hogerheijde}, {Folsom}, {Miotello}, {Fedele}, {Belloche}, {G{\"u}sten}, \& {Wyrowski}}]{2016Kama}
{Kama}, M., {Bruderer}, S., {van Dishoeck}, E.~F., {et~al.} 2016, \aap, 592, A83

\bibitem[{{Kama} {et~al.}(2015){Kama}, {Folsom}, \& {Pinilla}}]{2015Kama}
{Kama}, M., {Folsom}, C.~P., \& {Pinilla}, P. 2015, \aap, 582, L10

\bibitem[{{Kama} {et~al.}(2019){Kama}, {Shorttle}, {Jermyn}, {Folsom}, {Furuya}, {Bergin}, {Walsh}, \& {Keller}}]{2019Kama}
{Kama}, M., {Shorttle}, O., {Jermyn}, A.~S., {et~al.} 2019, \apj, 885, 114

\bibitem[{{Kasper} {et~al.}(2015){Kasper}, {Apai}, {Wagner}, \& {Robberto}}]{2015KasperHD110058ScatterLight}
{Kasper}, M., {Apai}, D., {Wagner}, K., \& {Robberto}, M. 2015, \apjl, 812, L33

\bibitem[{{Keyte} {et~al.}(2024){Keyte}, {Kama}, {Chuang}, {Cleeves}, {Drozdovskaya}, {Furuya}, {Rawlings}, \& {Shorttle}}]{2024Keyte}
{Keyte}, L., {Kama}, M., {Chuang}, K.-J., {et~al.} 2024, \mnras, 528, 388

\bibitem[{{Kilian}(1994)}]{1994Kilian}
{Kilian}, J. 1994, \aap, 282, 867

\bibitem[{{K{\'o}sp{\'a}l} {et~al.}(2013){K{\'o}sp{\'a}l}, {Mo{\'o}r}, {Juh{\'a}sz}, {{\'A}brah{\'a}m}, {Apai}, {Csengeri}, {Grady}, {Henning}, {Hughes}, {Kiss}, {Pascucci}, \& {Schmalzl}}]{2013KospalHD21997}
{K{\'o}sp{\'a}l}, {\'A}., {Mo{\'o}r}, A., {Juh{\'a}sz}, A., {et~al.} 2013, \apj, 776, 77

\bibitem[{{Kral} {et~al.}(2020{\natexlab{a}}){Kral}, {Davoult}, \& {Charnay}}]{2020Kralpla}
{Kral}, Q., {Davoult}, J., \& {Charnay}, B. 2020{\natexlab{a}}, Nature Astronomy, 4, 769

\bibitem[{{Kral} {et~al.}(2024){Kral}, {Huet}, {Bergez-Casalou}, {Th{\'e}bault}, {Charnoz}, \& {Fornasier}}]{2024Kral}
{Kral}, Q., {Huet}, P., {Bergez-Casalou}, C., {et~al.} 2024, \aap, 692, A70

\bibitem[{{Kral} {et~al.}(2019){Kral}, {Marino}, {Wyatt}, {Kama}, \& {Matr{\`a}}}]{2019KralHD131835}
{Kral}, Q., {Marino}, S., {Wyatt}, M.~C., {Kama}, M., \& {Matr{\`a}}, L. 2019, \mnras, 489, 3670

\bibitem[{{Kral} {et~al.}(2020{\natexlab{b}}){Kral}, {Matr{\`a}}, {Kennedy}, {Marino}, \& {Wyatt}}]{2020Kral}
{Kral}, Q., {Matr{\`a}}, L., {Kennedy}, G.~M., {Marino}, S., \& {Wyatt}, M.~C. 2020{\natexlab{b}}, \mnras, 497, 2811

\bibitem[{{Kral} {et~al.}(2017){Kral}, {Matr{\`a}}, {Wyatt}, \& {Kennedy}}]{Kraletal2017}
{Kral}, Q., {Matr{\`a}}, L., {Wyatt}, M.~C., \& {Kennedy}, G.~M. 2017, \mnras, 469, 521

\bibitem[{{Kunzli} {et~al.}(1997){Kunzli}, {North}, {Kurucz}, \& {Nicolet}}]{1997Geneva}
{Kunzli}, M., {North}, P., {Kurucz}, R.~L., \& {Nicolet}, B. 1997, \aaps, 122, 51

\bibitem[{Kupka {et~al.}(1999)Kupka, Piskunov, Ryabchikova, Stempels, \& Weiss}]{1999Kupka}
Kupka, Piskunov, N., Ryabchikova, T.~A., Stempels, H.~C., \& Weiss, W.~W. 1999, Astron. Astrophys. Suppl. Ser., 138, 119

\bibitem[{Kupka {et~al.}(2000)Kupka, Ryabchikova, Piskunov, Stempels, \& Weiss}]{2000Kupkaetal}
Kupka, F.~G., Ryabchikova, T.~A., Piskunov, N.~E., Stempels, H.~C., \& Weiss, W.~W. 2000, Open Astronomy, 9, 590

\bibitem[{{Kurucz}(1993)}]{1993Kurucz}
{Kurucz}, R. 1993, Robert Kurucz CD-ROM, 13

\bibitem[{{Landstreet}(1988)}]{1988Landstreet}
{Landstreet}, J.~D. 1988, \apj, 326, 967

\bibitem[{{Lieman-Sifry} {et~al.}(2016){Lieman-Sifry}, {Hughes}, {Carpenter}, {Gorti}, {Hales}, \& {Flaherty}}]{2016SifryHD138813HD110058ALMA}
{Lieman-Sifry}, J., {Hughes}, A.~M., {Carpenter}, J.~M., {et~al.} 2016, \apj, 828, 25

\bibitem[{{MacGregor} {et~al.}(2018){MacGregor}, {Weinberger}, {Hughes}, {Wilner}, {Currie}, {Debes}, {Donaldson}, {Redfield}, {Roberge}, \& {Schneider}}]{2018MacGregorHD32297}
{MacGregor}, M.~A., {Weinberger}, A.~J., {Hughes}, A.~M., {et~al.} 2018, \apj, 869, 75

\bibitem[{{Meeus} {et~al.}(2001){Meeus}, {Waters}, {Bouwman}, {van den Ancker}, {Waelkens}, \& {Malfait}}]{2001Meeus}
{Meeus}, G., {Waters}, L.~B.~F.~M., {Bouwman}, J., {et~al.} 2001, \aap, 365, 476

\bibitem[{{Mermilliod} {et~al.}(1997){Mermilliod}, {Mermilliod}, \& {Hauck}}]{1997Mermilliod}
{Mermilliod}, J.~C., {Mermilliod}, M., \& {Hauck}, B. 1997, \aaps, 124, 349

\bibitem[{{Meshkat} {et~al.}(2017){Meshkat}, {Mawet}, {Bryan}, {Hinkley}, {Bowler}, {Stapelfeldt}, {Batygin}, {Padgett}, {Morales}, {Serabyn}, {Christiaens}, {Brandt}, \& {Wahhaj}}]{2017Meshkat_age}
{Meshkat}, T., {Mawet}, D., {Bryan}, M.~L., {et~al.} 2017, \aj, 154, 245

\bibitem[{{Mittal} {et~al.}(2015){Mittal}, {Chen}, {Jang-Condell}, {Manoj}, {Sargent}, {Watson}, \& {Lisse}}]{2015Mittal_age}
{Mittal}, T., {Chen}, C.~H., {Jang-Condell}, H., {et~al.} 2015, \apj, 798, 87

\bibitem[{{Montgomery} \& {Welsh}(2012)}]{2012Montgomery49Ceti}
{Montgomery}, S.~L. \& {Welsh}, B.~Y. 2012, \pasp, 124, 1042

\bibitem[{{Moon} \& {Dworetsky}(1985)}]{1985Moon}
{Moon}, T.~T. \& {Dworetsky}, M.~M. 1985, \mnras, 217, 305

\bibitem[{{Mo{\'o}r} {et~al.}(2017){Mo{\'o}r}, {Cur{\'e}}, {K{\'o}sp{\'a}l}, {{\'A}brah{\'a}m}, {Csengeri}, {Eiroa}, {Gunawan}, {Henning}, {Hughes}, {Juh{\'a}sz}, {Pawellek}, \& {Wyatt}}]{2017Moor}
{Mo{\'o}r}, A., {Cur{\'e}}, M., {K{\'o}sp{\'a}l}, {\'A}., {et~al.} 2017, \apj, 849, 123

\bibitem[{{Mo{\'o}r} {et~al.}(2013){Mo{\'o}r}, {Juh{\'a}sz}, {K{\'o}sp{\'a}l}, {{\'A}brah{\'a}m}, {Apai}, {Csengeri}, {Grady}, {Henning}, {Hughes}, {Kiss}, {Pascucci}, {Schmalzl}, \& {Gab{\'a}nyi}}]{2013MoorHD21997}
{Mo{\'o}r}, A., {Juh{\'a}sz}, A., {K{\'o}sp{\'a}l}, {\'A}., {et~al.} 2013, \apjl, 777, L25

\bibitem[{{Morgan} {et~al.}(1943){Morgan}, {Keenan}, \& {Kellman}}]{Morganetal1943}
{Morgan}, W.~W., {Keenan}, P.~C., \& {Kellman}, E. 1943, {An atlas of stellar spectra, with an outline of spectral classification}

\bibitem[{{Napiwotzki} {et~al.}(1993){Napiwotzki}, {Schoenberner}, \& {Wenske}}]{1993Napiwotzki}
{Napiwotzki}, R., {Schoenberner}, D., \& {Wenske}, V. 1993, \aap, 268, 653

\bibitem[{{Nielsen} {et~al.}(2019){Nielsen}, {De Rosa}, {Macintosh}, {Wang}, {Ruffio}, {Chiang}, {Marley}, {Saumon}, {Savransky}, {Ammons}, {Bailey}, {Barman}, {Blain}, {Bulger}, {Burrows}, {Chilcote}, {Cotten}, {Czekala}, {Doyon}, {Duch{\^e}ne}, {Esposito}, {Fabrycky}, {Fitzgerald}, {Follette}, {Fortney}, {Gerard}, {Goodsell}, {Graham}, {Greenbaum}, {Hibon}, {Hinkley}, {Hirsch}, {Hom}, {Hung}, {Dawson}, {Ingraham}, {Kalas}, {Konopacky}, {Larkin}, {Lee}, {Lin}, {Maire}, {Marchis}, {Marois}, {Metchev}, {Millar-Blanchaer}, {Morzinski}, {Oppenheimer}, {Palmer}, {Patience}, {Perrin}, {Poyneer}, {Pueyo}, {Rafikov}, {Rajan}, {Rameau}, {Rantakyr{\"o}}, {Ren}, {Schneider}, {Sivaramakrishnan}, {Song}, {Soummer}, {Tallis}, {Thomas}, {Ward-Duong}, \& {Wolff}}]{2019Nielsen_age}
{Nielsen}, E.~L., {De Rosa}, R.~J., {Macintosh}, B., {et~al.} 2019, \aj, 158, 13

\bibitem[{{Nieva} \& {Przybilla}(2012)}]{2012Nievacosmicabun}
{Nieva}, M.~F. \& {Przybilla}, N. 2012, \aap, 539, A143

\bibitem[{{Olofsson} {et~al.}(2022){Olofsson}, {Th{\'e}bault}, {Kennedy}, \& {Bayo}}]{2022OlofssonHD32297}
{Olofsson}, J., {Th{\'e}bault}, P., {Kennedy}, G.~M., \& {Bayo}, A. 2022, \aap, 664, A122

\bibitem[{{Pace} {et~al.}(2006){Pace}, {Recio-Blanco}, {Piotto}, \& {Momany}}]{Pace2006vmic}
{Pace}, G., {Recio-Blanco}, A., {Piotto}, G., \& {Momany}, Y. 2006, \aap, 452, 493

\bibitem[{{Pearce} {et~al.}(2022){Pearce}, {Launhardt}, {Ostermann}, {Kennedy}, {Gennaro}, {Booth}, {Krivov}, {Cugno}, {Henning}, {Quirrenbach}, {Barcucci}, {Matthews}, {Ruh}, \& {Stone}}]{2022Pearce_age}
{Pearce}, T.~D., {Launhardt}, R., {Ostermann}, R., {et~al.} 2022, \aap, 659, A135

\bibitem[{{Pinilla} {et~al.}(2012){Pinilla}, {Benisty}, \& {Birnstiel}}]{2012Pinilla}
{Pinilla}, P., {Benisty}, M., \& {Birnstiel}, T. 2012, \aap, 545, A81

\bibitem[{Pinilla {et~al.}(2015)Pinilla, Birnstiel, \& Walsh}]{2015Pinilla}
Pinilla, P., Birnstiel, T., \& Walsh, C. 2015, \aap, 580, A105

\bibitem[{Piskunov {et~al.}(1995)Piskunov, Kupka, Ryabchikova, Weiss, \& Jeffery}]{1995Piskunovetal}
Piskunov, N.~E., Kupka, F., Ryabchikova, T.~A., Weiss, W.~W., \& Jeffery, C.~S. 1995, \aaps, 112, 525

\bibitem[{{Redfield}(2007)}]{2007RedfieldHd32297}
{Redfield}, S. 2007, \apjl, 656, L97

\bibitem[{{Rhee} {et~al.}(2007){Rhee}, {Song}, {Zuckerman}, \& {McElwain}}]{2007Rhee_age}
{Rhee}, J.~H., {Song}, I., {Zuckerman}, B., \& {McElwain}, M. 2007, \apj, 660, 1556

\bibitem[{{Ribas} {et~al.}(1997){Ribas}, {Jordi}, {Torra}, \& {Gimenez}}]{1997Ribas}
{Ribas}, I., {Jordi}, C., {Torra}, J., \& {Gimenez}, A. 1997, \aap, 327, 207

\bibitem[{{Rice} {et~al.}(2006){Rice}, {Armitage}, {Wood}, \& {Lodato}}]{2006Rice}
{Rice}, W.~K.~M., {Armitage}, P.~J., {Wood}, K., \& {Lodato}, G. 2006, \mnras, 373, 1619

\bibitem[{{Roberge} {et~al.}(2014){Roberge}, {Welsh}, {Kamp}, {Weinberger}, \& {Grady}}]{2014Roberge49Ceti}
{Roberge}, A., {Welsh}, B.~Y., {Kamp}, I., {Weinberger}, A.~J., \& {Grady}, C.~A. 2014, \apjl, 796, L11

\bibitem[{{R{\'o}{\.z}a{\'n}ski} {et~al.}(2022){R{\'o}{\.z}a{\'n}ski}, {Niemczura}, {Lemiesz}, {Posi{\l}ek}, \& {R{\'o}{\.z}a{\'n}ski}}]{2022suppnet}
{R{\'o}{\.z}a{\'n}ski}, T., {Niemczura}, E., {Lemiesz}, J., {Posi{\l}ek}, N., \& {R{\'o}{\.z}a{\'n}ski}, P. 2022, {SUPPNet: Spectrum normalization neural network}, Astrophysics Source Code Library, record ascl:2202.004

\bibitem[{Ryabchikova {et~al.}(2015)Ryabchikova, Piskunov, Kurucz, Stempels, Heiter, Pakhomov, \& Barklem}]{2015Ryabchikova}
Ryabchikova, T., Piskunov, N., Kurucz, R.~L., {et~al.} 2015, Physica Scripta, 90, 054005

\bibitem[{Ryabchikova {et~al.}(1997)Ryabchikova, Piskunov, Kupka, \& Weiss}]{1997Ryabchikovaetal}
Ryabchikova, T.~A., Piskunov, N.~E., Kupka, F., \& Weiss, W.~W. 1997, Open Astronomy, 6, 244

\bibitem[{{Saffe} {et~al.}(2021){Saffe}, {Miquelarena}, {Alacoria}, {Flores}, {Jaque Arancibia}, {Calvo}, {Mart{\'\i}n Girardi}, {Grosso}, \& {Collado}}]{2021Saffe}
{Saffe}, C., {Miquelarena}, P., {Alacoria}, J., {et~al.} 2021, \aap, 647, A49

\bibitem[{{Schlafly} \& {Finkbeiner}(2011)}]{SF2011}
{Schlafly}, E.~F. \& {Finkbeiner}, D.~P. 2011, \apj, 737, 103

\bibitem[{{Schlegel} {et~al.}(1998){Schlegel}, {Finkbeiner}, \& {Davis}}]{SFD1998}
{Schlegel}, D.~J., {Finkbeiner}, D.~P., \& {Davis}, M. 1998, \apj, 500, 525

\bibitem[{{Schneider} {et~al.}(2014){Schneider}, {Grady}, {Hines}, {Stark}, {Debes}, {Carson}, {Kuchner}, {Perrin}, {Weinberger}, {Wisniewski}, {Silverstone}, {Jang-Condell}, {Henning}, {Woodgate}, {Serabyn}, {Moro-Martin}, {Tamura}, {Hinz}, \& {Rodigas}}]{2014SchneiderHD32297}
{Schneider}, G., {Grady}, C.~A., {Hines}, D.~C., {et~al.} 2014, \aj, 148, 59

\bibitem[{Shulyak {et~al.}(2004)Shulyak, Tsymbal, Ryabchikova, St\"utz, \& Weiss}]{2004Shulyakllmodels}
Shulyak, D., Tsymbal, V., Ryabchikova, T., St\"utz, C., \& Weiss, W.~W. 2004, A\&A, 428, 993

\bibitem[{{Simpson} {et~al.}(2019){Simpson}, {Martell}, {Da Costa}, {Casey}, {Freeman}, {Horner}, {Ting}, {Nataf}, {Lewis}, {Ness}, {Zucker}, {Cottrell}, {{\v{C}}otar}, {Asplund}, {Bland-Hawthorn}, {Buder}, {D'Orazi}, {De Silva}, {Duong}, {Kos}, {Lin}, {Lind}, {Schlesinger}, {Sharma}, {Zwitter}, {Kafle}, \& {Nordlander}}]{2019SimpsonGALAH}
{Simpson}, J.~D., {Martell}, S.~L., {Da Costa}, G., {et~al.} 2019, \mnras, 482, 5302

\bibitem[{{Stasevic} {et~al.}(2023){Stasevic}, {Milli}, {Mazoyer}, {Lagrange}, {Bonnefoy}, {Faramaz-Gorka}, {M{\'e}nard}, {Boccaletti}, {Choquet}, {Shuai}, {Olofsson}, {Chomez}, {Ren}, {Rubini}, {Desgrange}, {Gratton}, {Chauvin}, {Vigan}, \& {Matthews}}]{Stasevic2023HD110058}
{Stasevic}, S., {Milli}, J., {Mazoyer}, J., {et~al.} 2023, \aap, 678, A8

\bibitem[{{Tanr{\i}verdi}(2013)}]{2013Tanriverdi}
{Tanr{\i}verdi}, T. 2013, \na, 25, 50

\bibitem[{{Tetzlaff} {et~al.}(2011){Tetzlaff}, {Neuh{\"a}user}, \& {Hohle}}]{2011Tetzlaff_age}
{Tetzlaff}, N., {Neuh{\"a}user}, R., \& {Hohle}, M.~M. 2011, \mnras, 410, 190

\bibitem[{{Turcotte}(2002)}]{2002Turcotte}
{Turcotte}, S. 2002, \apjl, 573, L129

\bibitem[{{Turon} {et~al.}(1993){Turon}, {Creze}, {Egret}, {Gomez}, {Grenon}, {Jahrei{\ss}}, {Requieme}, {Argue}, {Bec-Borsenberger}, {Dommanget}, {Mennessier}, {Arenou}, {Chareton}, {Crifo}, {Mermilliod}, {Morin}, {Nicolet}, {Nys}, {Prevot}, {Rousseau}, {Perryman}, {Arlot}, {Baglin}, {Barthes}, {Baylac}, {Brosche}, {Burnet}, {Delhaye}, {Dettbarn}, {Erbach}, {Figueras}, {Fricke}, {Helmer}, {Hemenway}, {Jordi}, {Lampens}, {Lederle}, {Lub}, {Manfroid}, {Mattci}, {Mazurier}, {Mermilliod}, {Morrison}, {Murray}, {Oblak}, {Perie}, {Pernier}, {Le Poole}, {Quijano}, {Rapaport}, {Sellier}, {Torra}, {Tucholke}, {de Vegt}, {Argyle}, {Bacchus}, {Baron}, {Calaf}, {Cordoni}, {Fabricius}, {Feaugas}, {Fehlberg}, {Florkowski}, {de Geus}, {Gibbs}, {Hartmann}, {Jauncey}, {Johnston}, {Marouard}, {Mekkas}, {Muinos}, {Nunez}, {Ochsenbein}, {de Orus}, {Paredes}, {Penston}, {Petersen}, {Peyrin}, {Robin}, {Roman}, {Rossello}, {Schwan}, {Sinachopoulos}, {White}, {Zacharias}, {Hog}, {Kovalevsky}, {van Leeuwen}, {Lindegren}, {Schutz},
  \& {Schrijver}}]{1993Turcon_sptype_vmag}
{Turon}, C., {Creze}, M., {Egret}, D., {et~al.} 1993, Bulletin d'Information du Centre de Donnees Stellaires, 43, 5

\bibitem[{{Ujjwal} {et~al.}(2020){Ujjwal}, {Kartha}, {Mathew}, {Manoj}, \& {Narang}}]{2020Ujjwal_age}
{Ujjwal}, K., {Kartha}, S.~S., {Mathew}, B., {Manoj}, P., \& {Narang}, M. 2020, \aj, 159, 166

\bibitem[{{Viana Almeida} {et~al.}(2009){Viana Almeida}, {Santos}, {Melo}, {Ammler-von Eiff}, {Torres}, {Quast}, {Gameiro}, \& {Sterzik}}]{2009VianaSACY}
{Viana Almeida}, P., {Santos}, N.~C., {Melo}, C., {et~al.} 2009, \aap, 501, 965

\bibitem[{{Wade, G. A.} {et~al.}(2001){Wade, G. A.}, {Bagnulo, S.}, {Kochukhov, O.}, {Landstreet, J. D.}, {Piskunov, N.}, \& {Stift, M. J.}}]{2001Wadeetal}
{Wade, G. A.}, {Bagnulo, S.}, {Kochukhov, O.}, {et~al.} 2001, \aap, 374, 265

\bibitem[{{Waters} \& {Waelkens}(1998)}]{WatersWaelkens1998}
{Waters}, L.~B.~F.~M. \& {Waelkens}, C. 1998, \araa, 36, 233

\bibitem[{{Youdin} \& {Goodman}(2005)}]{2005Youdin_and_Goodman}
{Youdin}, A.~N. \& {Goodman}, J. 2005, \apj, 620, 459

\bibitem[{{Zuckerman} \& {Song}(2012)}]{2012Zuckerman49Ceti}
{Zuckerman}, B. \& {Song}, I. 2012, \apj, 758, 77

\end{thebibliography}

\begin{appendix}
\onecolumn
\section{Additional Plots}
\begin{figure}[h]{\includegraphics[width=17cm, trim=0.0cm 0.2cm 0.0cm 0.0cm]{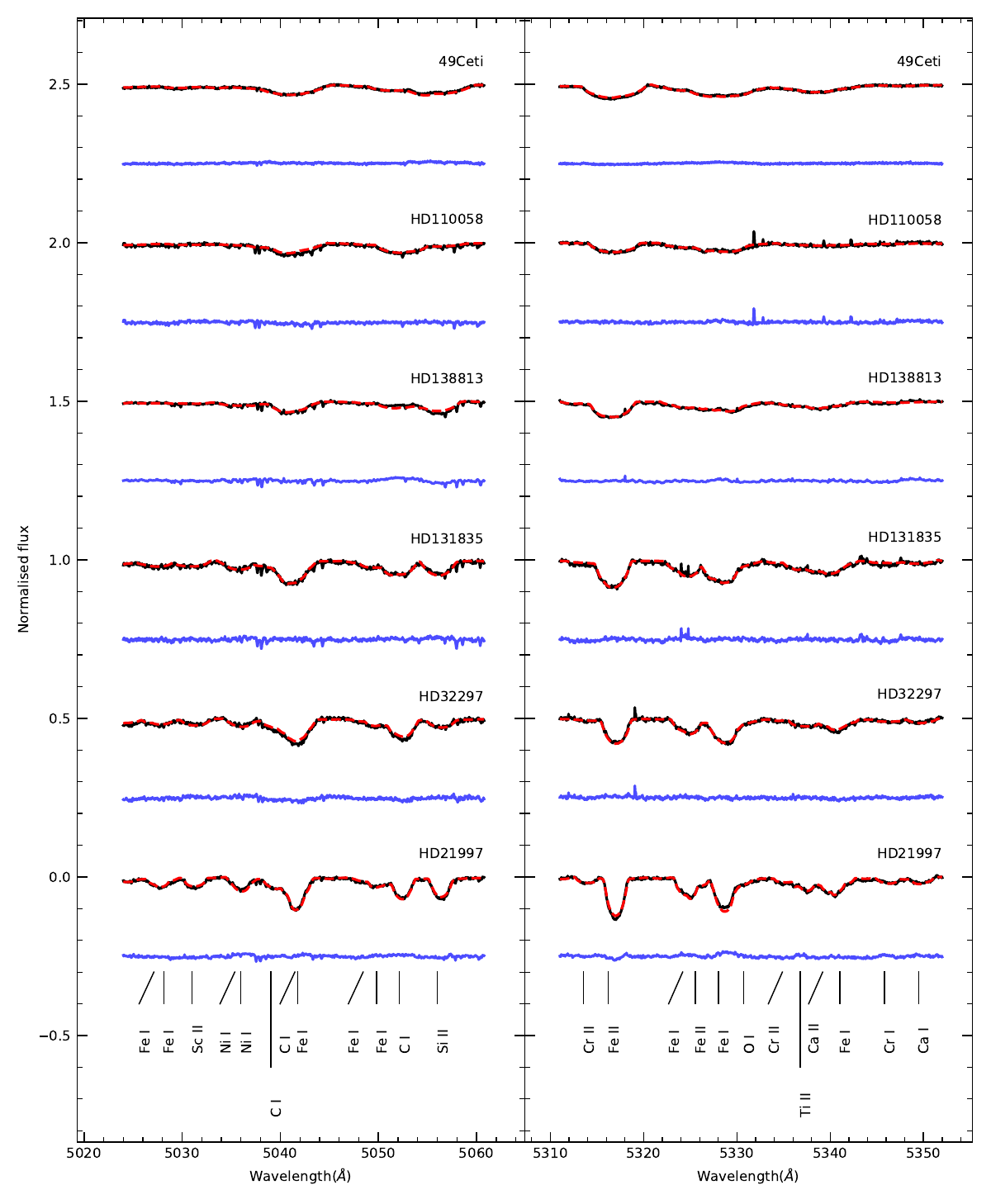}}
 \caption{Spectral fitting of the debris disk host star sample around C lines (left panel) and an O line (right panel). The spectra are shifted vertically for better visualisation. Black represents the observed spectrum, red represents the best-fit synthetic spectrum, and blue represents the difference between the observed and synthetic spectrum. The spectra are sorted in increasing order of \vsini\ value from bottom to top.}
 \label{fig:spectral_fitting_C_and_O}
\end{figure}

\newpage
\section{Additional Data}

\begin{table*}[h]
\renewcommand{\arraystretch}{1.3}
\small
    \caption{Details about the spectra of the six stars used for this study.}
    \centering
    \begin{tabular}{l   l   c   c  c}
    \hline
    \hline
    \multicolumn{1}{c}{Star} & \multicolumn{1}{c}{Instrument}& \multicolumn{1}{c}{Observation Time} & \multicolumn{1}{c}{Exposure Time} &\multicolumn{1}{c}{Combined S/N}\\
    & & & (seconds)&(at $\sim$5800$\AA$ per spectral bin)\\
    \hline
    \starone& FEROS&2017-09-24T05:22:54&360 & 479\\
    & &2017-09-24T08:10:48&360 &\\
    \hline
    \startwo & ESPADONS& 2005-08-25T11:34:35 & 200& 634\\
    & &2005-08-25T11:38:41&200 &\\
    & &2005-08-25T11:42:49&200 &\\
    & &2005-08-25T11:46:56 &200  &\\
    \hline
    \starthree& FEROS&2016-03-29T03:57:24&1800 &336\\
    & &2016-03-29T08:05:12&1800 &\\
    & &2016-03-26T03:46:21&1200 &\\
    & &2016-03-26T07:24:06&1800 &\\
    & &2016-03-27T05:13:19&1800 &\\
    & &2016-03-27T07:23:14&1800 &\\
    & &2016-03-28T02:44:03&1800 &\\
    & &2016-03-28T04:24:05&1800 &\\
    & &2016-03-28T07:36:57&1800 &\\
    & &2016-03-29T02:32:29&1800 &\\
    \hline
    \starfour& FEROS&2016-03-29T04:44:33&1200 &561\\
    & &2016-03-29T08:39:38&1200 &\\
    & &2016-03-26T04:10:54&700  &\\
    & &2016-03-26T07:58:38&1000 &\\
    & &2016-03-27T03:23:44&660 &\\
    & &2016-03-27T06:04:29&1200 &\\
    & &2016-03-27T07:57:17&1800 &\\
    & &2016-03-28T03:59:59&1200 &\\
    & &2016-03-28T08:12:05&1200 &\\
    \hline
    \starfive& FEROS&2017-04-02T04:40:17 & 900 &382\\
    & & 2017-04-03T04:14:38 & 900 &\\
    & & 2017-04-04T04:02:12 & 1400 &\\
    & & 2017-04-05T03:29:40 & 1200 &\\
    & & 2017-04-06T03:31:39 & 1200 &\\
    & & 2017-04-09T04:10:00 & 900 &\\
    \hline
    \starsix & FEROS & 2017-09-24T07:26:15 & 2100 &349\\
    & & 2017-09-26T07:04:12 & 2100 &\\
    & & 2017-09-26T08:47:26 & 2100 &\\
    & & 2017-09-30T07:05:12 & 2100 &\\
    & & 2017-10-01T06:51:33 & 2100 &\\
    \hline
  \end{tabular}
  \label{tab:stellar_spectra}
\end{table*}

\begin{table}[h]
\renewcommand{\arraystretch}{1.3}
\caption{Reference sample of protoplanetary disk-hosting stars for comparison.}
\centering

\begin{tabular}{l l l l l l l}
\hline
\hline
System & Age & \multicolumn{1}{c}{C} & \multicolumn{1}{c}{O} & \multicolumn{1}{c}{Fe} & \multicolumn{1}{c}{$\rm log_{10}($\mdot$)$} & \multicolumn{1}{c}{Meeus}\\
\hline
$\rm HD\,31648$ & 6.2 & -3.63 ± 0.07 & -3.24 ± 0.05 & -4.43 ± 0.13 & -6.9 & IIa \\
$\rm HD\,163296$ & 4.1 & -3.78 ± 0.25 & -3.27 ± 0.15 & -4.35 ± 0.15 & -7.16 & IIa \\
$\rm HD\,244604$ & 1.9 & -3.65 ± 0.17 & -3.19 ± 0.08 & -4.31 ± 0.24 & -7.2 & IIa \\
$\rm HD\,190073$ & 1.6 & -3.68 ± 0.14 & -3.22 ± 0.11 & -4.38 ± 0.06 & -8.68 & IIa \\
$\rm HD\,104237$ & 2.06 & -3.58 ± 0.20 & -3.17 ± 0.15 & -4.39 ± 0.18 & -7.45 & IIa \\
$\rm HD\,142666$ & 6.9 & -3.58 ± 0.15 & -3.14 ± 0.15 & -4.80 ± 0.11 & -7.77 & IIa \\
$\rm HD\,144432$ & 5.5 & -3.75 ± 0.18 & -3.13 ± 0.10 & -4.66 ± 0.09 & -7.74 & IIa \\
$\rm HD\,101412$ & 1.2 & -3.54 ± 0.12 & -3.08 ± 0.09 & -5.04 ± 0.19 & -7.04 & IIa/Ia \\
$\rm HD\,31293$ & 2.6 & -3.29 ± 0.22 & -3.23 ± 0.20 & -4.87 ± 0.22 & -7.74 & Ia \\
$\rm HD\,36112$ & 1.5 & -3.57 ± 0.16 & -3.14 ± 0.10 & -4.45 ± 0.14 & -6.05 & Ia \\
$\rm HD\,179218$ & 1.1 & -3.41 ± 0.16 & -3.06 ± 0.13 & -4.99 ± 0.13 & -6.72 & Ia \\
$\rm HD\,100546$ & 10 & -3.68 ± 0.20 & -3.25 ± 0.20 & -5.67 ± 0.08 & -7.23 & Ia \\
$\rm HD\,250550$ & 0.25 &\multicolumn{1}{c}{ - }&\multicolumn{1}{c}{ - }& -5.36 ± 0.12 & -7.8 & Ia \\
$\rm HD\,142527$ & 0.86 &\multicolumn{1}{c}{ - }&\multicolumn{1}{c}{ - }& -4.59 & -7.16 & Ia \\
$\rm HD\,139614$ & 13.5 & -3.71 ± 0.22 & -3.29 ± 0.10 & -5.03 ± 0.13 & -7.99 & Ia \\
$\rm HD\,34282$ & 7.81 &\multicolumn{1}{c}{ - }&\multicolumn{1}{c}{ - }& -5.3 ± 0.1 & -7.71 & Ib \\
$\rm HD\,141569$ & 4 & -3.59 ± 0.29 & -3.01 ± 0.10 & -5.21 ± 0.32 & -8.37 & Ib \\
$\rm HD\,169142$ & 8.5 & -3.51 ± 0.12 & -3.34 ± 0.13 & -5.09 ± 0.11 & -7.4 & Ib \\
$\rm HD\,100453$ & 10 & -3.44 ± 0.09 &\multicolumn{1}{c}{ - }& -4.57 ± 0.19 & -8.04 & Ib \\
$\rm HD\,135344\,B$ & 8.93 &\multicolumn{1}{c}{ - }&\multicolumn{1}{c}{ - }& -4.56 ± 0.10 & -7.69 & Ib \\
$\rm HD\,245185$ & 5.5 & -3.68 ± 0.15 & -3.13 ± 0.17 & -5.23 ± 0.33 & -7.2 & I \\
\hline
\end{tabular}
\tablefoot{The stellar abundances are taken from \citet[][and references therein]{2015Kama}. The Meeus column identifies the specific Meeus group \citep{2001Meeus} to which the star belongs, with the "a" and "b" sub-classifications based on the presence or absence of emission features from warm, small silicate grains, respectively.}
\label{tab:Protoplanetary disk ref abundance}
\end{table}

\begin{table}[h]
\renewcommand{\arraystretch}{1.3}
\caption{Stellar abundances of Sco-Cen association members (Table \ref{tab:sco-cen_stars}). The references to the abundances for all the stars are listed in Table \ref{tab:Sco-Cen detailed abundance3}.}
\centering
\small
\begin{tabular}{l l l l l l l l l}
\hline
\hline
Star & \multicolumn{1}{c}{C} & \multicolumn{1}{c}{N} & \multicolumn{1}{c}{O} & \multicolumn{1}{c}{Na} & \multicolumn{1}{c}{Mg} & \multicolumn{1}{c}{Al} & \multicolumn{1}{c}{Si} & \multicolumn{1}{c}{S} \\
\hline
\multicolumn{9}{c}{T < 6250 K} \\
\hline
CP-52 5025 &  &  &  &  &  &  & -4.63 ± 0.10 &  \\
CD-51 6900 &  &  &  &  &  &  & -4.68 ± 0.08 &  \\
CD-36 10208 &  &  &  &  &  &  & -4.90 ± 0.11 &  \\
CD-49 4947 &  &  &  &  &  &  & -4.64 ± 0.10 &  \\
CD-23 13281 &  &  &  &  &  &  & -4.52 ± 0.10 &  \\
CD-39 10162 &  &  &  &  &  &  & -4.63 ± 0.09 &  \\
CD-51 10295 &  &  &  &  &  &  & -4.58 ± 0.08 &  \\
CD-22 11502 &  &  &  &  &  &  & -4.69 ± 0.07 &  \\
TYC 9034 968 1 &  &  &  &  &  &  & -4.63 ± 0.11 &  \\
CD-40 7581 &  &  &  &  &  &  & -4.58 ± 0.11 &  \\
TYC7886 1894 1 &  &  &  &  &  &  & -4.65 ± 0.08 &  \\
CD-25 11330 &  &  &  &  &  &  & -4.67 ± 0.06 &  \\
CD-50 6815 & -3.67 &  & -1.72 & -5.91 & -4.59 & -5.48 &  &  \\
CD-34 10180 &  &  &  &  &  &  & -4.68 ± 0.09 &  \\
CD-40 8031 &  &  &  &  &  &  & -4.59 ± 0.09 &  \\
CD-49 7027 & -3.82 &  & -2.85 & -5.93 & -4.57 & -5.45 &  &  \\
CD-31 13486 &  &  &  &  &  &  & -4.53 ± 0.04 &  \\
HD 102458 & -3.84 &  & -2.25 & -5.91 & -4.51 & -5.63 & -4.51 &  \\
\hline
\multicolumn{9}{c}{T > 6250 K} \\
\hline
HD 112810 & -4.11 &  & -3.56 & -6.21 & -4.82 & -6.03 & -4.86 &  \\
HD 135344 & -3.28 &  &  &  &  &  & -4.50 ± 0.10 &  \\
HD 120326 & -3.74 ± 0.09 &  &  &  & -4.52 ± 0.15 & -6.06 ± 0.14 & -4.17 ± 0.13 &  \\
HD 133803 & -3.64 ± 0.10 &  &  &  & -4.27 ± 0.21 & -5.78 ± 0.22 & -4.22 ± 0.25 &  \\
HD 144432 & -3.75 ± 0.18 &  & -3.13 ± 0.10 & -5.82 ± 0.09 & -4.42 ± 0.04 &  & -4.77 ± 0.24 & -4.78 ± 0.05 \\
HD 100453 & -3.44 ± 0.09 & -4.12 &  & -5.53 ± 0.14 & -4.43 ± 0.09 &  & -4.32 ± 0.07 & -4.90 ± 0.07 \\
HD 142666 & -3.58 ± 0.15 &  & -3.14 ± 0.15 & -5.73 ± 0.15 & -4.57 ± 0.06 &  & -4.87 ± 0.20 & -4.66 ± 0.15 \\
HD 115820 & -3.7 &  &  &  & -4.32 & -6.01 & -4.28 &  \\
HD 95086 & -3.69 &  &  &  & -4.23 & -6 & -4.54 &  \\
HD 139614 & -3.71 ± 0.22 & -4.13 ± 0.13 & -3.29 ± 0.1 & -6.1 ± 0.12 & -4.7 ± 0.07 &  & -5.06 ± 0.13 & -5.14 ± 0.15 \\
Eta Leo & -3.975 ± 0.08 & -3.555 ± 0.19 & -3.4 ± 0.14 &  & -4.645 ± 0.07 & -6.02 ± 0.03 & -4.55 ± 0.18 & -4.99 ± 0.2 \\
HD 122980 & -3.68 ± 0.09 & -4.24 ± 0.08 & -3.28 ± 0.06 &  & -4.5 ± 0.05 &  &  &  \\
HR 4898 & -3.91 ± 0.06 & -4.51 ± 0.15 & -3.62 ± 0.18 &  & -4.98 ± 0.01 & -6.25 ± 0.15 & -5.27 ± 0.09 &  \\
HD 120324 &  & -4.22 ± 0.12 &  &  &  &  &  &  \\
HR 5285v & -3.82 ± 0.17 & -4.52 ± 0.13 & -3.73 ± 0.16 &  & -4.67 ± 0.01 & -6.19 ± 0.13 & -5.17 ± 0.11 &  \\
HR 4853 & -3.95 ± 0.05 & -4.39 ± 0.11 & -3.47 ± 0.16 &  & -4.61 ± 0.01 & -5.90 ± 0.06 & -4.66 ± 0.05 &  \\
HR 6165 & -3.62 ± 0.10 & -3.91 ± 0.17 & -3.41 ± 0.14 &  & -4.34 ± 0.01 & -5.64 ± 0.21 & -4.43 ± 0.13 &  \\
HD 149438 & -3.7 ± 0.12 & -3.84 ± 0.12 & -3.23 ± 0.08 &  & -4.38 ± 0.03 &  & -4.48 ± 0.06 &  \\
\hline
\end{tabular}
\label{tab:Sco-Cen detailed abundance1}
\end{table}

\begin{table}[h]
\renewcommand{\arraystretch}{1.3}
\caption{Stellar abundances of Sco-Cen association members (Table \ref{tab:sco-cen_stars}).}
\centering
\small
\begin{tabular}{l l l l l l l l l}
\hline
\hline
Star & \multicolumn{1}{c}{Ca} & \multicolumn{1}{c}{Sc} & \multicolumn{1}{c}{Ti} & \multicolumn{1}{c}{V} & \multicolumn{1}{c}{Cr} & \multicolumn{1}{c}{Mn} & \multicolumn{1}{c}{Fe} & \multicolumn{1}{c}{Ni} \\
\hline
\multicolumn{9}{c}{T < 6250 K} \\
\hline
CP-52 5025 &  &  &  &  &  &  & -4.57 ± 0.07 & -5.68 ± 0.09 \\
CD-51 6900 &  &  &  &  &  &  & -4.65 ± 0.08 & -5.75 ± 0.11 \\
CD-36 10208 &  &  &  &  &  &  & -4.77 ± 0.07 & -5.71 ± 0.13 \\
CD-49 4947 &  &  &  &  &  &  & -4.58 ± 0.06 & -5.86 ± 0.13 \\
CD-23 13281 &  &  &  &  &  &  & -4.42 ± 0.17 & -5.71 ± 0.16 \\
CD-39 10162 &  &  &  &  &  &  & -4.64 ± 0.08 & -5.81 ± 0.10 \\
CD-51 10295 &  &  &  &  &  &  & -4.58 ± 0.05 & -5.78 ± 0.10 \\
CD-22 11502 &  &  &  &  &  &  & -4.68 ± 0.07 & -5.74 ± 0.11 \\
TYC 9034 968 1 &  &  &  &  &  &  & -4.61 ± 0.11 & -5.88 ± 0.10 \\
CD-40 7581 &  &  &  &  &  &  & -4.49 ± 0.07 & -5.71 ± 0.12 \\
TYC7886 1894 1 &  &  &  &  &  &  & -4.68 ± 0.07 & -5.69 ± 0.16 \\
CD-25 11330 &  &  &  &  &  &  & -4.66 ± 0.07 & -5.80 ± 0.11 \\
CD-50 6815 & -5.54 & -8.75 & -7.02 & -7.88 & -6.22 & -6.41 &  & -5.55 \\
CD-34 10180 &  &  &  &  &  &  & -4.61 ± 0.07 & -5.68 ± 0.12 \\
CD-40 8031 &  &  &  &  &  &  & -4.53 ± 0.07 & -5.71 ± 0.09 \\
CD-49 7027 & -5.54 & -8.56 & -7.06 & -7.68 & -6.16 & -6.45 &  & -5.55 \\
CD-31 13486 &  &  &  &  &  &  & -4.49 ± 0.06 & -5.71 ± 0.09 \\
HD 102458 & -5.81 & -8.65 & -7.2 & -8.01 & -6.35 & -6.62 &  & -5.79 \\
\hline
\multicolumn{9}{c}{T > 6250 K} \\
\hline
HD 112810 & -5.96 & -9.33 & -7.37 & -8.34 & -6.92 & -7.02 &  & -6.2 \\
HD 135344 & -5.64 ± 0.27 & -9.26 & -7.46 ± 0.01 &  & -6.27 ± 0.06 &  & -4.55 ± 0.10 & -5.67 ± 0.19 \\
HD 120326 & -5.56 ± 0.17 & -8.79 ± 0.28 & -7.10 ± 0.14 &  & -6.47 ± 0.08 & -6.90 ± 0.14 & -4.64 ± 0.10 &  \\
HD 133803 & -5.48 ± 0.36 & -8.63 ± 0.31 & -6.95 ± 0.15 &  & -6.17 ± 0.11 & -6.57 ± 0.14 & -4.48 ± 0.15 &  \\
HD 144432 & -5.74 ± 0.15 & -9.05 ± 0.15 & -7.22 ± 0.16 & -8.26 ± 0.40 & -6.47 ± 0.23 & -6.92 ± 0.15 & -4.66 ± 0.09 & -5.96 ± 0.15 \\
HD 100453 & -5.61 ± 0.12 & -9.11 ± 0.03 & -7.15 ± 0.18 & -7.72 & -6.30 ± 0.10 & -6.37 ± 0.17 & -4.55 ± 0.19 & -5.78 ± 0.15 \\
HD 142666 & -5.94 ± 0.18 & -9.19 ± 0.23 & -7.35 ± 0.20 & -8.36 ± 0.40 & -6.56 ± 0.21 & -7.04 ± 0.15 & -4.80 ± 0.11 & -6.12 ± 0.13 \\
HD 115820 & -5.73 & -8.73 & -6.94 &  & -6.36 & -6.64 & -4.56 & -5.99 \\
HD 95086 & -5.71 & -8.97 & -7.12 &  & -6.41 & -6.75 & -4.58 &  \\
HD 139614 & -6.14 ± 0.16 & -9.35 ± 0.15 & -7.47 ± 0.14 & -8.36 ± 0.5 & -6.76 ± 0.11 & -7.39 ± 0.19 & -5.03 ± 0.13 & -6.29 ± 0.14 \\
Eta Leo & -5.7 ± 0.08 & -9.41 ± 0.21 & -7.5 ± 0.19 & -8.37 ± 0.14 & -6.405 ± 0.14 & -6.63 ± 0.2 & -4.5 ± 0.2 & -5.8 ± 0.19 \\
HD 122980 &  &  &  &  &  &  & -4.56 ± 0.11 &  \\
HR 4898 &  &  &  &  &  &  &  &  \\
HD 120324 &  &  &  &  &  &  &  &  \\
HR 5285v &  &  &  &  &  &  &  &  \\
HR 4853 &  &  &  &  &  &  &  &  \\
HR 6165 &  &  &  &  &  &  &  &  \\
HD 149438 &  &  &  &  &  &  & -4.46 ± 0.09 &  \\
\hline
\end{tabular}
\label{tab:Sco-Cen detailed abundance2}
\end{table}

\begin{table}[h]
\renewcommand{\arraystretch}{1.3}
\caption{Stellar abundances of Sco-Cen association members (Table \ref{tab:sco-cen_stars}).}
\centering
\small
\begin{tabular}{l l l l l l l c}
\hline
\hline
Star & \multicolumn{1}{c}{Cu} & \multicolumn{1}{c}{Zn} & \multicolumn{1}{c}{Sr} & \multicolumn{1}{c}{Y} & \multicolumn{1}{c}{Ba} & \multicolumn{1}{c}{La} & Reference \\
\hline
\multicolumn{8}{c}{T < 6250 K} \\
\hline
CP-52 5025 &  &  &  &  &  &  & 9 \\
CD-51 6900 &  &  &  &  &  &  & 9 \\
CD-36 10208 &  &  &  &  &  &  & 9 \\
CD-49 4947 &  &  &  &  &  &  & 9 \\
CD-23 13281 &  &  &  &  &  &  & 9 \\
CD-39 10162 &  &  &  &  &  &  & 9 \\
CD-51 10295 &  &  &  &  &  &  & 9 \\
CD-22 11502 &  &  &  &  &  &  & 9 \\
TYC 9034 968 1 &  &  &  &  &  &  & 9 \\
CD-40 7581 &  &  &  &  &  &  & 9 \\
TYC7886 1894 1 &  &  &  &  &  &  & 9 \\
CD-25 11330 &  &  &  &  &  &  & 9 \\
CD-50 6815 & -8.05 & -7.14 &  & -8.9 & -9.12 & -10.64 & 10 \\
CD-34 10180 &  &  &  &  &  &  & 9 \\
CD-40 8031 &  &  &  &  &  &  & 9 \\
CD-49 7027 & -7.97 & -7.32 &  & -9.06 & -9.27 & -10.72 & 10 \\
CD-31 13486 &  &  &  &  &  &  & 9 \\
HD 102458 & -8.09 & -7.44 &  & -9.95 & -9.07 & -10.94 & 1 \\
\hline
\multicolumn{8}{c}{T > 6250 K} \\
\hline
HD 112810 & -8.00 & -8.22 &  & -10.42 & -10.27 & -10.89 & 1 \\
HD 135344 &  & -7.85 &  &  &  &  & 2 \\
HD 120326 &  &  & -9.01 ± 0.22 & -9.72 ± 0.14 & -9.66 ± 0.23 &  & 3 \\
HD 133803 &  &  & -8.97 ± 0.33 & -9.69 ± 0.21 & -9.30 ± 0.27 &  & 3 \\
HD 144432 &  & -7.53 ± 0.20 & -9.36 ± 0.30 & -9.76 ± 0.13 & -9.65 ± 0.16 & -11.16 ± 0.40 & 4 \\
HD 100453 &  & -7.92 &  &  & -9.89 &  & 2 \\
HD 142666 &  & -7.86 ± 0.30 & -9.52 ± 0.19 & -9.90 ± 0.22 & -9.77 ± 0.18 &  & 4 \\
HD 115820 &  &  & -9.04 & -9.62 & -9.33 &  & 3 \\
HD 95086 &  &  & -8.78 & -9.86 & -9.57 &  & 3 \\
HD 139614 & -8.56 ± 0.4 & -8.26 ± 0.3 &  & -10.19 ± 0.15 & -10.25 ± 0.21 &  & 4 \\
Eta Leo &  &  & -9.84 ± 0.05 & -9.67 ± 0.15 & -9.99 & -8.53 ± 0.06 & 5 \\
HD 122980 &  &  &  &  &  &  & 6 \\
HR 4898 &  &  &  &  &  &  & 7 \\
HD 120324 &  &  &  &  &  &  & 8 \\
HR 5285v &  &  &  &  &  &  & 7 \\
HR 4853 &  &  &  &  &  &  & 7 \\
HR 6165 &  &  &  &  &  &  & 7 \\
HD 149438 &  &  &  &  &  &  & 6 \\
\hline
\end{tabular}
\tablebib{(1) \citet{2018BuderGALAHDR2}, (2) \citet{2004AckeWaelkens}, (3) \citet{2021Saffe}, (4) \citet{2012Folsom}, (5) \citet{2013Tanriverdi}, (6) \citet{2012Nievacosmicabun}, (7) \citet{1994Kilian}, (8) \citet{2017AhmedSigut}, (9) \citet{2009VianaSACY}, (10) \citet{2019SimpsonGALAH}.}
\label{tab:Sco-Cen detailed abundance3}
\end{table}
\end{appendix}
\end{document}